\documentclass[twocolumn,showpacs,preprintnumbers,amsmath,amssymb,showkeys]{revtex4-2}
\usepackage{graphicx}
\usepackage{dcolumn}
\usepackage{bm}

\begin{document}

\title{Optical properties of anisotropic Dirac semimetals}
 \author{I. Kup\v{c}i\'{c} and J. Kordi\'{c}}     
 \affiliation{
   Department of Physics, Faculty of Science, University of Zagreb, 
   P.O. Box 331,  HR-10002 Zagreb,  Croatia}

\begin{abstract} 
The current-dipole conductivity formula for doped three-dimensional Dirac semimetals is derived by using a modified gauge-invariant tight-binding approach.
    In a heavily doped regime, the effective number of charge carriers $n_{\alpha \alpha}^{\rm eff}$ in the Drude contribution is found to be by a factor of 4 larger than the nominal electron concentration $n$.
    However, its structure is the same as in standard Fermi liquid theory.
    In a lightly doped regime, on the other hand, the ratio $n_{\alpha \alpha}^{\rm eff}/n$ is much larger, with much more complex structure of $n_{\alpha \alpha}^{\rm eff}$.
    It is shown that the dc resistivity and reflectivity date measured in two TlBiSSe samples can be easily understood, 
even in the relaxation-time approximation, provided that finite quasiparticle lifetime effects in the momentum distribution functions are properly taken into account.

\end{abstract}
\maketitle

\section{Introduction}
Over the past fifteen years,
there has been a lot of experimental \cite{Novak15,LeMardele23,Martino19}
and theoretical interest \cite{Armitage18,Tabert16,Kotov16,Zhang09,Liu10,Singh12,Neupane14}
in transport and electrodynamic properties of three-dimensional (3D) Dirac semimetal phases.
    Such a phase is usually found in the phase diagram at the critical point of a topological phase transition between a normal 
insulator and a topological insulator.
    In TlBi(S$_{1-x}$Se$_x$)$_2$ there is the phase transition between the normal insulator TlBiS$_2$ and the topological 
insulator TlBiSe$_2$.
    This system is particularly interesting because by tuning the ratio of Tl:Bi during synthesis different Dirac semimetal samples 
can be obtained characterized by different values of the nominal electron concentration $n$.
    In addition, the properties of samples with $n \lesssim 10^{17}$ cm$^{-3}$ are found to be strongly influenced by the synthesis quality.

Reconciling anomalous transport and electrodynamic properties of such 3D systems with a linear band dispersion with the theory based on a simple anisotropic 3D Dirac model is the subject of the present paper.
    Here we use the multiband current-dipole Kubo approach to determine the structure of the dynamical conductivity tensor $\sigma_{\alpha \alpha}(\omega)$.
    The exact form of electron-photon coupling functions is obtained by using a modified version of the common tight-binding minimal substitution. \cite{Kupcic07,Kupcic14}
    All damping effects are taken into account phenomenologically by using two different (intraband and interband) relaxation rates and one quasiparticle lifetime.
    This approach is found to be successful in explaining seemingly inconsistent properties of ultraclean and dirty lightly doped 
graphene samples. \cite{Kupcic16}
    The present numerical results show that for the doping level $n \approx 10^{19}$ cm$^{-3}$ (heavily doped regime) the effects of the finite quasiparticle lifetime on $\sigma_{\alpha \alpha}(\omega)$ 
can be safely neglected.
    This approximation leads to the common textbook multiband conductivity formula \cite{Wooten72,Platzman73,Ziman79,Ando02}
in which the momentum distribution functions are replaced by the corresponding Fermi-Dirac distribution functions.
    On the other hand, for $n \lesssim 10^{17}$ cm$^{-3}$ (lightly doped regime), a finite quasiparticle lifetime, taken into account in the way consistent with the Ward identity relations, \cite{Mahan90,Kupcic16,Kupcic15}
leads to different results depending upon whether the sample is dirty or clean.

The paper is organized as follows.
	In Sec. II, we use lightly doped graphene as an example to explain the difference between the nominal concentration 
of charge carriers $n$ and the effective number of charge carriers $n_{\alpha \alpha}^{\rm eff}$, as well as to show how the ratio between these two numbers relates to two different expressions for the electron mobility.
    In Sec.~III, the Bloch energies and the Bloch functions of the anisotropic 3D Dirac model are calculated by using transformation 
in which the effects of a finite Dirac mass are separated from the dependence on other model parameters.
    A modified version of the tight-binding minimal substitution from Appendix A is applied to the anisotropic ordinary Drude model 
in Sec.~IV.
    This section also includes the structure of intraband and interband current vertices as well as the final analytical expression 
for the current-dipole conductivity formula.
    In Sec.~V, the numerical results for the real part of the dynamical conductivity are presented.
    The dynamical conductivity tensor is found to be
a result of a complicated interplay among four energy scales: the Fermi energy, the Dirac mass parameter, 
intraband and interband damping energies, and temperature.
	The lightly doped regime with the Dirac mass not too large is found to be  particularly interesting because 
in this case the threshold energy for interband electron-hole excitations becomes comparable with the damping energies and/or with $k_{\rm B} T$, resulting in a complicated structure of both the dynamical conductivity and the dc conductivity.
    In Sec.~VI, the relation between the reciprocal effective-mass tensor and the cyclotron mass is briefly discussed.
    Section VII contains concluding remarks.

\section{DC conductivity of lightly doped graphene}
In the relaxation-time approximation, the dc conductivity of a general multiband model can be represented by the following current-dipole dc conductivity formula 
[the $\omega =0$ form of the multiband dynamical conductivity tensor $\sigma_{\alpha \alpha} (\omega)$ from Sec.~IV] \cite{Kupcic16}
\begin{eqnarray}
&& \hspace{-5mm} 
\sigma_{\alpha \alpha}^{\rm dc} =
\sum_{LL'} \frac{1}{V} \sum_{{\bf k} \sigma}
\frac{\hbar |J^{LL'}_{\alpha}({\bf k},{\bf k}_+)|^2}{\varepsilon_{LL'}({\bf k},{\bf k}_+)}[n_{L'}({\bf k}_+)-n_L({\bf k})]
\nonumber \\
&& \hspace{5mm}
\times \frac{\hbar \Gamma^{LL'}_\alpha({\bf k})}{\varepsilon_{LL'}^2({\bf k},{\bf k}_+) + [\hbar \Gamma^{LL'}_\alpha({\bf k})]^2},
\label{eq1}
\end{eqnarray}
where ${\bf k}_+ = {\bf k} +{\bf q}$.
    Here, the $J^{LL'}_{\alpha}({\bf k},{\bf k}_+)$ are the intraband ($L=L'$) and interband ($L\neq L'$) current vertex functions, 
    the $\varepsilon_{LL'}({\bf k},{\bf k}_+) =\varepsilon_{L}({\bf k})-\varepsilon_{L'}({\bf k}_+)$ are the renormalized electron-hole pair energies.
    The band dispersions $\varepsilon_{L}({\bf k})$ and the current vertex functions are usually described by simple theoretical models \cite{Kupcic07,Kupcic14}
or by using different {\it ab initio} methods \cite{Despoja13}.
    The $\Gamma^{LL'}_\alpha({\bf k})$ are the intraband and interband relaxation rates which represent the imaginary parts of the corresponding electron-hole self-energies.
    In the leading approximation, they can be treated as phenomenological parameters which are obtained by fitting measured resistivity and reflectivity data.
    The momentum distribution functions $n_L({\bf k})$ enable us to
treat clean and dirty electronic systems on an equal footing. \cite{Carbotte10,Kupcic16,Tabert16}
    In clean systems, $n_L({\bf k}) \approx f_{L} ({\bf k})$, where $f_{L} ({\bf k}) = f(\varepsilon_{L} ({\bf k}))$ is the Fermi-Dirac distribution function.
    
The sum $\sum_{\bf k}$ runs over the first Brillouin zone, and the band index $L$ runs over all bands in question.
    In graphene, for example, the band index $L$ runs over two bands, and in the spinless fermion representation for conduction electrons in Dirac semimetals, 
the sum $\sum_\sigma$ is missing and the sum $\sum_L$ runs over four bands.
    When the band dispersions are simplified by using the Dirac cone approximation, then the sum $\sum_{\bf k}$ is restricted to a small region around the Dirac points in which the dispersions are nearly linear in wave vector.
    Hereafter, this restricted sum will be labeled by $\sum_{\bf k}^*$.

\subsection{Electron mobility $\mu$ in graphene}
It is generally agreed that in pristine graphene the intraband and interband contributions to the dc conductivity $\sigma_{\alpha \alpha}^{\rm dc}$ are equally important.
    In this case, the dc conductivity can be shown in the following two equivalent ways, 
\cite{Kupcic16}
\begin{eqnarray}
&& \hspace{-10mm} 
\sigma_{\alpha \alpha}^{\rm dc} = \frac{e^2}{m\Gamma_{1\alpha}} n^{\rm eff}_{\alpha \alpha} = \frac{e^2}{m\Gamma_{1\alpha}} \left(n^{\rm intra}_{\alpha \alpha}+n^{\rm inter,1}_{\alpha \alpha}\right),
\label{eq2}
\end{eqnarray}
or
\begin{eqnarray}
&& \hspace{-10mm} 
\sigma_{\alpha \alpha}^{\rm dc} = \frac{e^2}{m\Gamma_{1\alpha}} n^{\rm intra}_{\alpha \alpha} +\frac{e^2}{m\Gamma_{2\alpha}} n^{\rm inter,2}_{\alpha \alpha}.
\label{eq3}
\end{eqnarray}
    Here \cite{Ashcroft76}
\begin{eqnarray}
&& \hspace{-5mm} n^{\rm intra}_{\alpha \alpha} = \frac{1}{V} \sum_{L{\bf k} \sigma} m 
[v_\alpha^{L} ({\bf k})]^2
\bigg( -\frac{\partial n_L({\bf k})}{\partial \varepsilon_L({\bf k})}\bigg)\frac{\Gamma_{1\alpha}}{\Gamma^{LL}_\alpha({\bf k})}
\label{eq4}
\end{eqnarray}
is the intraband part of the total number of charge carriers that participate in the dc conductivity, and $\Gamma_{1\alpha}$ is the intraband relaxation rate averaged over the Fermi surface
[in the leading approximation, $\Gamma_{1\alpha}/\Gamma^{LL}_\alpha({\bf k})$ can be replaced by 1]. 
    As discussed in detail below, this effective number of charge carriers 
must not be confused with the nominal concentration of charge carriers $n$.

Two expressions for the interband effective number of charge carriers in Eqs.~(\ref{eq2}) and (\ref{eq3}) are given by
\begin{eqnarray}
&& \hspace{-7mm} n^{{\rm inter,}i}_{\alpha \alpha} = 
\frac{1}{V} \sum_{L \neq L'}\sum_{ {\bf k} \sigma}
\frac{m}{e^2} |J^{LL'}_{\alpha}({\bf k},{\bf k}_+)|^2
\nonumber \\
&& \hspace{0mm} \times  \frac{n_{L'}({\bf k}_+)-n_L({\bf k})}{
\varepsilon_{LL'}({\bf k},{\bf k}_+)}
\frac{\hbar^2 \Gamma^{LL'}_\alpha({\bf k})\Gamma_{i\alpha}}{\varepsilon_{LL'}^2({\bf k},{\bf k}_+) + (\hbar \Gamma^{LL'}_\alpha({\bf k}))^2},
\label{eq5}
\end{eqnarray}
where $i = 1, 2$.
    For $\Gamma_{1\alpha} =\Gamma_{2\alpha}$, we obtain $n^{{\rm inter,}1}_{\alpha \alpha} = n^{{\rm inter,}2}_{\alpha \alpha} \equiv n^{{\rm inter}}_{\alpha \alpha}$.
    In heavily doped graphene, the interband contribution to $\sigma_{\alpha \alpha}^{\rm dc}$ is found to be negligible, 
resulting in
\begin{eqnarray}
&& \hspace{-10mm} 
 n^{\rm eff}_{\alpha \alpha} \approx n^{\rm intra}_{\alpha \alpha}.
\label{eq6} 
\end{eqnarray}

In 3D Dirac systems, the dc conductivity is given by the same expressions.
    According to Eq.~(\ref{eq2}), it can be understood as a function of two temperature-dependent factors, the effective number of charge carriers and the averaged relaxation time $\tau_\alpha = 1/\Gamma_{1\alpha}$,
    In the anisotropic case, the anisotropy in $\sigma_{\alpha \alpha}^{\rm dc}$ originates from the anisotropy in both
$\tau_\alpha$ and $n^{\rm eff}_{\alpha \alpha}$.

In experimental analyses, it is usual to define the electron mobility $\mu^{\rm exp}_\alpha$ in the following way \cite{Ziman79,Novoselov05,Zhang05,Novak15}
\begin{eqnarray}
&& \hspace{-10mm} 
\sigma_{\alpha \alpha}^{\rm dc} = e \mu^{\rm exp}_\alpha n.
\label{eq7} 
\end{eqnarray}
    This electron mobility is nothing but
the dc conductivity shown in units of mobility in the case where the effective number of charge carriers $n^{\rm eff}_{\alpha \alpha}$ is replaced by the nominal concentration of charge carriers $n$. 
    In theory, on the other hand, it is more convenient to define the electron mobility $\mu^{\rm th}_\alpha$ as the product of the electron relaxation time $1/\Gamma_{1\alpha}$ and the factor $e/m$, resulting in the relation \cite{Kupcic16}
\begin{eqnarray}
&& \hspace{-10mm} 
\sigma_{\alpha \alpha}^{\rm dc} = e \mu^{\rm th}_\alpha n^{\rm eff}_{\alpha \alpha},
\label{eq8} 
\end{eqnarray}
and 
\begin{eqnarray}
&& \hspace{-10mm} 
 \mu^{\rm exp}_\alpha = \frac{n^{\rm eff}_{\alpha \alpha}}{n}\mu^{\rm th}_\alpha.
\label{eq9} 
\end{eqnarray}
    For electrons with parabolic dispersion, these two expressions for $\mu$ represent essentially the same physical quantity, 
the mobility of all electrons that participate in the dc conductivity (namely, $n^{\rm eff}_{\alpha \alpha} \approx n$ in this case).
    However, for two-dimensional (2D) and 3D Dirac electrons, the effective number of charge carriers is not simply related to $n$.
    In the isotropic 3D case, for example, one obtains $n \propto k_{\rm F}^3$ and $n^{\rm intra}_{\alpha \alpha} \propto k_{\rm F}^2$.
    Therefore, the ratio $\mu^{\rm exp} /\mu^{\rm th}$ is proportional to $n^{-1/3}$.
    Similarly, in the 2D Dirac case the relation is of the form $\mu^{\rm exp} /\mu^{\rm th} \propto 1/\sqrt{n}$.
    A brief discussion of the relation between $n^{\rm eff}_{\alpha \alpha}$ and $n$ in 3D Dirac semimetals is given in Sec. VI.

\begin{figure}
  \centerline{\includegraphics[width=17pc]{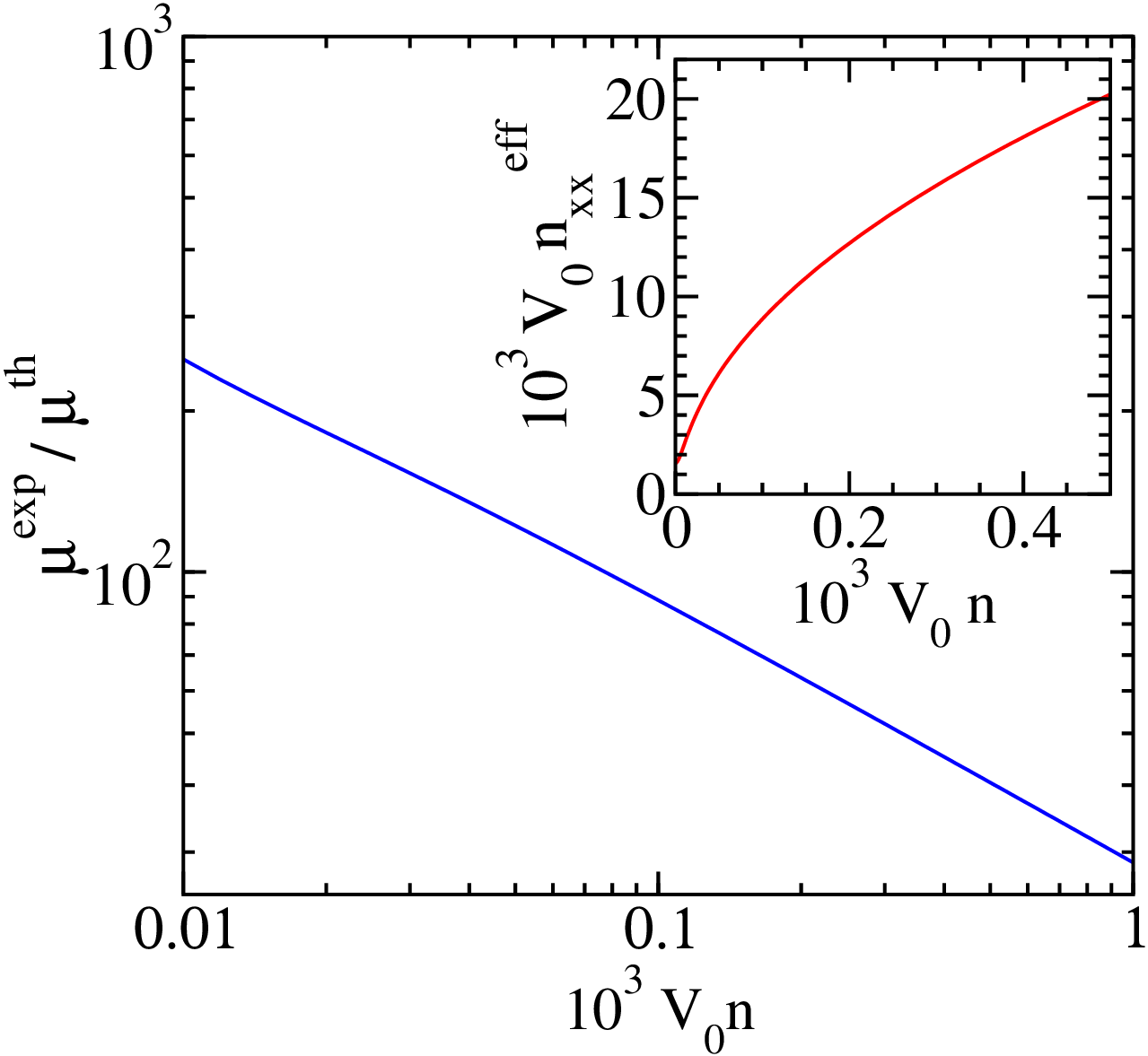}}
   \caption{(Color online) Main figure: the doping dependence of $\mu^{\rm exp}/\mu^{\rm th}$ in lightly doped graphene calculated by using Eq.~(\ref{eq1}),
   for $n_L({\bf k}) = f_{L} ({\bf k})$, $\hbar \Gamma_{1\alpha} = \hbar \Gamma_{2\alpha}=0.5$  meV, and $T= 60$ K.
   Inset: the doping dependence of $n^{\rm eff}_{\alpha \alpha}$ in the same case.
   $V_0 = \sqrt{3}a^2/2$ is the primitive cell volume.
}
  \end{figure}  
    
Figure 1 shows the dependence of $\mu^{\rm exp}/\mu^{\rm th}$ on $n$ in clean lightly doped graphene
calculated by using Eq.~(\ref{eq1}).
    The inset of figure shows the doping dependence of $n^{\rm eff}_{\alpha \alpha}$ calculated at $T = 60$ K and the solid line in Fig.~2 is the same function calculated at zero temperature.
    In Fig.~1, it should be noticed that in the lightly doped regime $\mu^{\rm exp}$ is much larger than $\mu^{\rm th}$.
    Therefore, caution is in order regarding
theoretical explanation of measured mobility data in different 2D and 3D Dirac systems.
    
\begin{figure}
   \centerline{\includegraphics[width=17pc]{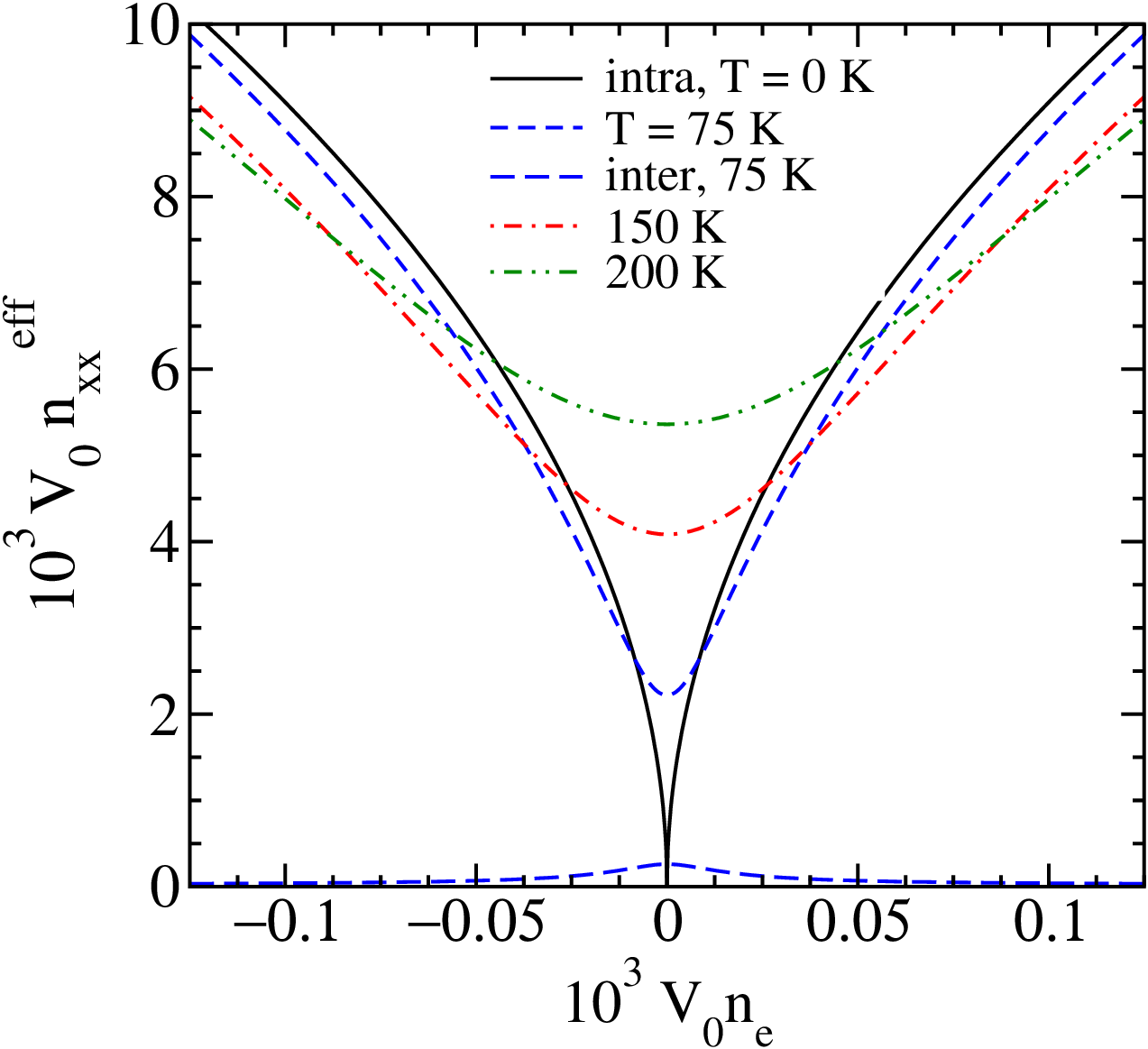}}
   \caption{(Color online)  The doping dependence of $n^{\rm eff}_{\alpha \alpha}$ in lightly doped graphene at $T = 75$, 150, and 200 K, for 
   $\Gamma^{LL'}_\alpha({\bf k})/\Gamma_{LL'}=1$, $\hbar \Gamma_1 = \hbar \Gamma_2 = 5$ meV and $n_L ({\bf k}) = f_L ({\bf k})$.
   The solid line is $n^{\rm intra}_{\alpha \alpha}$ calculated at $T=0$ K.
   The nominal concentration of charge carriers is $n=|n_{\rm e}|$.
}
  \end{figure} 	
  
\subsection{Thermally activated conduction electrons}
Figure 2 shows the doping dependence of $n^{\rm eff}_{\alpha \alpha}$ in clean lightly doped graphene at different temperatures.
    $n^{\rm intra}_{\alpha \alpha}$ calculated at zero temperature, $n^{\rm intra,0}_{\alpha \alpha}$, is also shown 
[$V_0 n^{\rm intra,0}_{\alpha \alpha}\approx 0.91 (m/m_{xx}) \sqrt{V_0 n}$, solid line].
    Not surprisingly, in pristine graphene $n^{\rm intra,0}_{\alpha \alpha}$ vanishes, simply because the Fermi surface comprises only two Dirac points.
    Therefore, the corresponding dc conductivity, $\sigma_{\alpha \alpha}^{\rm dc,0}$, originates solely from the $\omega = 0$ interband processes.
    In the current-dipole conductivity formula, the number $n^{\rm eff,0}_{\alpha \alpha} = n^{\rm inter,0}_{\alpha \alpha}$ multiplied by $e \mu^{\rm th}_\alpha$ gives the well-known result
$\sigma_{\alpha \alpha}^{\rm dc,0}= (\pi^2 e/2 \hbar)$.  \cite{Li08,Ziegler07,Carbotte10}

In pristine graphene, the thermally activated intraband contribution to $n^{\rm eff}_{\alpha \alpha}$ becomes dominant even for not too high temperatures
(compare $n^{\rm eff}_{\alpha \alpha}$ with $n^{\rm inter}_{\alpha \alpha}$ both calculated at $T= 75$ K).
    This contribution decreases with increasing doping, and
at each temperature, there is a critical value of $n$, $n_{\rm c}(T)$, at which the difference $n^{\rm eff}_{\alpha \alpha} - n^{\rm intra,0}_{\alpha \alpha}$  changes the sign.
    Such a complicated temperature dependence can be easily understood if we compare the alternative form of Eq.~(\ref{eq4}) 
with the usual expression for $n$.
    When integrated by parts with respect to ${\bf k}$
for $\Gamma^{LL}_\alpha({\bf k})= \Gamma_{1}$, Eq.~(\ref{eq4}) leads to
\cite{Ashcroft76,Kupcic14}
\begin{eqnarray}
&& \hspace{-5mm} n^{\rm intra}_{\alpha \alpha} = \frac{1}{V} \sum_{L{\bf k} \sigma} 
\gamma_{\alpha\alpha}^{LL} ({\bf k}) f_L({\bf k}).
\label{eq10}
\end{eqnarray}
    $n$ is given by Eq.~({\ref{eq10}) as well, with the reciprocal effective mass tensor $\gamma_{\alpha\alpha}^{LL} ({\bf k}) = m \partial^2 \varepsilon_L ({\bf k}) /\partial p_\alpha^2$ replaced by unity ($p_\alpha = \hbar k_\alpha$).
    In graphene, 
$\gamma_{\alpha\alpha}^{LL} ({\bf k}) = m \partial^2 \varepsilon_L ({\bf k}) /\partial p_\alpha^2$ strongly depends on the wave vector, and, consequently,
the interplay between the temperature dependence of $\beta$ and the temperature dependence of chemical potential in Eq.~({\ref{eq10}) leads to a small decrease in $n^{\rm intra}_{\alpha \alpha}$ for $n> n_{\rm c}(T)$  and to pronounced thermal effects for $n < n_{\rm c}(T)$.

For further considerations of the 3D Dirac model it is appropriate also to show 
the expression (\ref{eq10}) in the form valid in the Dirac cone approximation.
    In this case, the electrons in upper bands are shown in the electron picture and the holes in lower bands in the hole picture, resulting in
\begin{eqnarray}
&& \hspace{-5mm} n^{\rm intra}_{\alpha \alpha} = \frac{1}{V} {\sum_{L{\bf k} \sigma}}^* 
s_L \gamma_{\alpha\alpha}^{LL} ({\bf k}) f(s_L \varepsilon_L({\bf k})).
\label{eq11}
\end{eqnarray}
    Here the sign $s_L = {\rm sign} (\varepsilon_L({\bf k}))$ is equal to $+1$ and $-1$
(see the related discussion of $n^{\rm intra}_{\alpha \alpha}$ in Appendix D).

\section{Anisotropic 3D Dirac model}
It is well known that the salient low-energy features of the Dirac semimetals can be captured
by using a simple effective model in which the conduction electrons are described by the generalized anisotropic 
3D Dirac model with a finite Dirac mass.
	In the spinless fermion representation the bare Hamiltonian is given by the $4 \times 4$ effective Hamiltonian \cite{Zhang09,Liu10}
\begin{eqnarray}
&& \hspace{-5mm}
H_0 =  \sum_{ll'{\bf k} } \big[ \varepsilon_0({\bf k})\delta_{l,l'} +  H_0^{ll'} ({\bf k}) \big]c^\dagger_{ l{\bf k}} c_{l'{\bf k}},
\label{eq12}
\end{eqnarray}
where
\begin{eqnarray}
&& \hspace{-5mm}
\left(  H_0^{ll'} ({\bf k}) \right) 
=  \left( \begin{array}{cccc} 
{\cal M}({\bf k}) & 0 & K_z & K_- \\
0 & {\cal M}({\bf k}) & K_+ & -K_z \\
K_z & K_- & - {\cal M}({\bf k}) & 0 \\
K_+ & -K_z & 0 & - {\cal M}({\bf k})
\end{array} \right) 
\nonumber \\
&& \hspace{11mm}
\equiv
\left( \begin{array}{cccc} 
{\cal M}({\bf k}) {\rm I}_2\ & {\bf K} \cdot {\bm \sigma} \\
{\bf K} \cdot {\bm \sigma} & - {\cal M}({\bf k}) {\rm I}_2
\end{array} \right) \equiv  H_{ll}
\label{eq13}
\end{eqnarray}
and ${\cal M}({\bf k}) = M - \Delta M ({\bf k})$.
	Here ${\rm I}_2$ is the identity matrix of size 2, the $\sigma_\alpha$ are three Pauli matrices,
$M$ is the Dirac mass parameter, $K_\alpha = v_{{\rm F} \alpha} p_\alpha$, $K_{\pm} = K_x \pm i K_y$,
and the $v_{{\rm F} \alpha}$ are the corresponding Fermi velocities.
    The model parameters can be obtained by fitting the band structure of different {\it ab initio} calculations \cite{Zhang09,Singh12,Neupane14,LeMardele23}.
    In the leading approximation, the dispersive corrections $\varepsilon_0({\bf k})$ and $\Delta M ({\bf k})$ 
can be set to zero.

In Eq.~(\ref{eq12}) the molecular orbital index $l$ describes two different orbitals (labeled here by $a$ and $b$) 
with two values of the angular momentum ($j = \pm 1/2$). 
    We use the representation of delocalized molecular orbitals $\{ l {\bf k} \}$, where $|l {\bf k} \rangle = c^\dagger_{ l{\bf k}} |0 \rangle$,
    $l = A, B, C, D$, $|A {\bf k} \rangle \equiv |a {\bf k} 1/2 \rangle$, $|B {\bf k} \rangle \equiv |a {\bf k} -1/2 \rangle$,
$|C {\bf k} \rangle \equiv |b {\bf k} 1/2 \rangle$, and $|D {\bf k} \rangle \equiv |b {\bf k} -1/2 \rangle$.

In order to separate the features related to a finite Dirac mass from the dependence of the Bloch energies and the Bloch functions on other model parameters, we introduce an auxiliary representation of electronic states 
which will be called here 
the $\{ m {\bf k} \}$ representation. 
	The vectors 
\begin{eqnarray}
&& \hspace{-10mm}
|m{\bf k}\rangle \equiv c^\dagger_{ m{\bf k}} |0 \rangle = \sum_l  U^*_{\bf k}(m,l) |l {\bf k} \rangle, 
\label{eq14}
\end{eqnarray}
$m =a, b, c, d$, are given in terms of elements of the transformation matrix 
\begin{eqnarray}
&& \hspace{-10mm}
\hat U_{\bf k}^{lm} =
\big(  U_{\bf k}(l,m) \big) 
=  \left( \begin{array}{cccc} 
\sqrt{2}/2 & \sqrt{2}/2 & 0 & 0 \\
\sqrt{2}/2  &  -\sqrt{2}/2  & 0 & 0 \\
0 & 0 & u_{\bf k} & v_{\bf k} \\
0 & 0 & - v^*_{\bf k}& u^*_{\bf k}
\end{array} \right) .
\label{eq15}
\end{eqnarray}
    Here
\begin{eqnarray}
&& \hspace{-5mm}
u_{\bf k} = \frac{K_z + K_+}{\sqrt{2}K}, \hspace{3mm} v_{\bf k} = \frac{K_z - K_+}{\sqrt{2}K},
\label{eq16}
\end{eqnarray}
and $K^2 = \sum_\alpha K_\alpha^2$.
	In this representation the  bare Hamiltonian becomes
\begin{eqnarray}
&& \hspace{-5mm}
H_0 =  \sum_{mm'{\bf k} } \big[ \varepsilon_0({\bf k})\delta_{m,m'} + H_0^{mm'} ({\bf k})\big] c^\dagger_{ m{\bf k}} c_{m'{\bf k}},
\label{eq17}
\end{eqnarray}
where
\begin{eqnarray}
&& \hspace{-10mm} 
\left(  H_0^{mm'} ({\bf k}) \right) 
=  
\left( \begin{array}{cccc} 
{\cal M}({\bf k}) {\rm I}_2 & K {\rm I}_2 \\
K {\rm I}_2 & - {\cal M}({\bf k}) {\rm I}_2
\end{array} \right).
\label{eq18}
\end{eqnarray}
	Notice that the states with $m =a$ and $c$ are decoupled from the states with $m=b$ and $d$, as well as that the matrix elements $U_{\bf k}(l,m)$ are independent of ${\cal M}({\bf k})$.
		
The solutions to the Schr\"{o}dinger equation 
\begin{eqnarray}
&& \hspace{-5mm}
\sum_{m'}\big[ H_0^{mm'} ({\bf k}) - \big(\varepsilon_L({\bf k}) -\varepsilon_0({\bf k}) \big)\delta_{m,m'}\big] U^*_{\bf k}(L,m') = 0
\nonumber \\
\label{eq19}
\end{eqnarray}
are the Bloch energies $\varepsilon_L ({\bf k})$ and the transformation matrix elements $U_{\bf k}(m,L)$.
    A straightforward calculation gives the bare Hamiltonian   
 \begin{eqnarray}
&& \hspace{-5mm}
H_0 =  \sum_{L {\bf k} } \varepsilon_L ({\bf k}) c^\dagger_{ L{\bf k}} c_{L{\bf k}},
\label{eq20}
\end{eqnarray}
with four bands with the dispersions
\begin{eqnarray}
&& \hspace{-5mm}
\varepsilon_L ({\bf k})  -\varepsilon_0({\bf k}) = s_L \varepsilon ({\bf k}) = s_L \sqrt{K^2 + {\cal M}^2({\bf k})},
\label{eq21}
\end{eqnarray}
where $L = 1, 2, 3, 4$ is the band index, and $s_1 = s_2 = - s_3 = -s_4 = 1$.
    The second transformation matrix is given by
\begin{eqnarray}
\hat U_{\bf k}^{mL} =
\left(  U_{\bf k}(m,L) \right) 
=  \left( \begin{array}{cccc} 
U_{\bf k}{\rm I}_2  &  V_{\bf k} {\rm I}_2 \\
V_{\bf k}{\rm I}_2   & -U_{\bf k} {\rm I}_2
\end{array} \right),
\label{eq22}
\end{eqnarray}
with
\begin{eqnarray}
&& \hspace{-5mm}
	U_{\bf k} = \cos \frac{\phi ({\bf k})}{2}, \hspace{3mm} V_{\bf k} = \sin \frac{\phi ({\bf k})}{2}.
\label{eq23}
\end{eqnarray}
	The auxiliary phase $\phi ({\bf k})$ is given in the usual way,
\begin{eqnarray}
&& \hspace{-5mm}	
\tan \phi ({\bf k}) = \frac{K}{{\cal M}({\bf k})}.
\label{eq24}
\end{eqnarray}
    Finally, the total transformation matrix, $\hat U_{\bf k}$ between the delocalized molecular orbitals $|l{\bf k}\rangle$ 
and the Bloch states $|L{\bf k}\rangle = c^\dagger_{ L{\bf k}} |0 \rangle$, can be shown in the following form 
\begin{eqnarray}
&& \hspace{-10mm}
\hat U_{\bf k} = \left( U_{\bf k}(l,L)\right) = \hat U_{\bf k}^{lm} \hat U_{\bf k}^{mL},
\label{eq25}
\end{eqnarray}
where
\begin{eqnarray}
&& \hspace{-10mm}
U_{\bf k}(l,L) = \sum_m  U_{\bf k}(l,m) U_{\bf k}(m,L).
\label{eq26}
\end{eqnarray}

In the following, we take $\varepsilon_0({\bf k}) = \Delta M ({\bf k}) = 0$ as an example
(to be referred to as the anisotropic ordinary 3D Dirac model).
	In this case, the band structure comprises two bands with 
the dispersion $\varepsilon_1 ({\bf k}) = \varepsilon_2 ({\bf k}) = \varepsilon ({\bf k}) = \sqrt{K^2+ M^2}$ 
and two bands with the dispersion $\varepsilon_3 ({\bf k}) = \varepsilon_4 ({\bf k}) = -\varepsilon ({\bf k})$.
	For $M=0$, the dispersions are $\varepsilon ({\bf k}) = \sqrt{K^2}$ and 
$-\varepsilon ({\bf k})$, respectively, and the auxiliary phase $\phi ({\bf k})$ is equal to $\pi/2$, resulting in $U_{\bf k} = V_{\bf k} = \sqrt{2}/2$.

\section{Dynamical conductivity tensor in general spinless multiband models}
According to Ref.~\cite{Zhang09}, in multiband electronic systems with inversion symmetry and with a finite spin-orbit coupling, it is convenient to use the spinless fermion representation for conduction electrons.
	In this case, the orbital index $l$ runs over all molecular orbitals in the primitive cell 
which participate in building bands under consideration.
	The total Hamiltonian $\widetilde H_0$ from Appendix A, which descibes the coupling between electrons and external electromagnetic fields, can be shown in the following form 
\begin{eqnarray}
&& \hspace{-5mm}
\widetilde H_0 =   \sum_{ll'}\sum_{n n'} c^\dagger_{ln } \delta_{{\bf R}_n,{\bf R}_{n'}} 
H_0^{ll'}(\hat {\bf p}_{n'}- \frac{e}{c} {\bf A} ({\bf R}_{n}))c_{l'n'}.
\label{eq27}
\end{eqnarray}

In the absence of static magnetic fields, we perform the Taylor expansion in the vector potential of $\widetilde{H}_0$ to the second order.
	The result is
\begin{eqnarray}
&& \hspace{-5mm}
\widetilde H_0 = H_0 + H^{\rm ext}_1 + H^{\rm ext}_2 + \ldots.
\label{eq28}
\end{eqnarray}
    The resulting bare Hamiltonian and the related coupling Hamiltonian shown in the Bloch representation
are given, respectively, by
\begin{eqnarray}
&& \hspace{-10mm}
H_0 = \sum_{L{\bf k}} \varepsilon_L ({\bf k}) c^\dagger_{L {\bf k}} c_{L {\bf k}}
\label{eq29}
\end{eqnarray}
and
\begin{eqnarray}
 && \hspace{-12mm}
H^{\rm ext}_1  = -\frac{1}{c} \sum_{{\bf q} \alpha} A_{\alpha} ({\bf q})
\hat J_{\alpha} (-{\bf q}),
\nonumber \\
&& \hspace{-12mm}
H^{\rm ext}_2 =  \frac{e^2}{2mc^2} \sum_{{\bf q} {\bf q}'\alpha \beta} 
A_{\alpha} ({\bf q}-{\bf q}')A_{\beta} ({\bf q}') \hat  \gamma_{\alpha \beta}(-{\bf q};2).
\label{eq30}
 \end{eqnarray} 
	In the coupling Hamiltonian, the current density operator and the bare diamagnetic density operator are given by
\begin{eqnarray}
&& \hspace{-10mm}
\hat J_{\alpha} ({\bf q}) = \sum_{LL'}\sum_{{\bf k}} J_{\alpha}^{LL'} ({\bf k}, {\bf k}_+)
c^\dagger_{L{\bf k}} c_{L'{\bf k}+{\bf q}},
\label{eq31} \\
&& \hspace{-10mm}
\hat \gamma_{\alpha \beta}({\bf q};2) = \sum_{LL'}\sum_{{\bf k}} \gamma_{\alpha \beta}^{LL'} ({\bf k},{\bf k}_+;2)
c^\dagger_{L{\bf k}} c_{L'{\bf k}+{\bf q}}.
\label{eq32}
\end{eqnarray}
	 Here $J_\alpha^{LL'} ({\bf k}, {\bf k}_+)$ and 
$\gamma_{\alpha \beta}^{LL'} ({\bf k},{\bf k}_+;2)$ are the corresponding bare vertex functions.
    The current vertices $J_\alpha^{LL'} ({\bf k}, {\bf k}_+)$ are given by Eq.~(\ref{eqB6}) in Appendix B, and the $\gamma_{\alpha \beta}^{LL'} ({\bf k},{\bf k}_+;2)$ are given by a similar expression.
    Notice that in the ordinary Dirac model Eq.~(\ref{eqB4}) gives 
$\gamma_{\alpha \beta}^{LL} ({\bf k},{\bf k}_+;2)=0$, resulting in $H^{\rm ext}_2=0$.
    As discussed in Appendix D in more detail, the vertex functions $\gamma_{\alpha \beta}^{LL} ({\bf k},{\bf k}_+;2)=0$ play an important role in understanding the transverse conductivity sum rule.

The dynamical conductivity tensor of such a multiband model can be represented by the current-dipole Kubo formula.
	In the relaxation-time approximation, this formula reads as \cite{Kupcic14}
\begin{eqnarray}
&& \hspace{-10mm} 
\sigma_{\alpha \alpha} ({\bf q}, \omega) =
\sum_{LL'} \frac{1}{V} \sum_{{\bf k}}
\frac{{\it i} \hbar |J^{LL'}_{\alpha}({\bf k},{\bf k}_+)|^2}{\varepsilon_{LL'}({\bf k},{\bf k}_+)}
\nonumber \\
&& \hspace{5mm}
\times \frac{n_{L'}({\bf k}_+)-n_L({\bf k})}{\hbar \omega + \varepsilon_{LL'}({\bf k},{\bf k}_+) + {\it i}\hbar \Gamma^{LL'}_\alpha ({\bf k})}.
\label{eq33}
\end{eqnarray}
    For $n_L({\bf k})=f_L({\bf k})$, this espression represents
a compact way of writing
Eqs.~(\ref{eq41}) and (\ref{eq43}).

\subsection{Ordinary 3D Dirac model}
In the anisotropic ordinary 3D Dirac model, the current vertex functions $J_{\alpha}^{LL'} ({\bf k}, {\bf k}_+) \approx J_{\alpha}^{LL'} ({\bf k})$ 
are given by Eqs.~(\ref{eqB9})$-$(\ref{eqB11}) in Appendix B.
	As expected, the intraband current vertices satisfy the electron-group-velocity theorem \cite{Kittel87}
\begin{equation}
J_{\alpha}^{LL} ({\bf k}) = s_L e v_{{\rm F}\alpha}  \frac{K_\alpha}{\sqrt{K^2 + M^2}}
\equiv e v_{\alpha}^{L} ({\bf k}) = e \frac{\partial \varepsilon_L ({\bf k})}{\partial p_\alpha}.
\label{eq34} 
\end{equation}
	Here $v_{\alpha}^{L} ({\bf k})$ is the electron group velocity in the band labeled by the band index $L$.
	Another general conclusion is that the interband current vertices between the bands that are degenerate 
in energy are equal to zero, i.e.
\begin{equation}
J_{\alpha}^{12} ({\bf k}) = J_{\alpha}^{21} ({\bf k}) = J_{\alpha}^{34} ({\bf k}) = J_{\alpha}^{43} ({\bf k}) = 0.
\label{eq35} 
\end{equation}
	A direct consequence of the result (\ref{eq35}) is the fact  that
the coherence factors in the interband conductivity tensor can be shown in terms of the coherence factors 
\begin{equation}
J_{\alpha}^{-+} ({\bf k}) J_{\beta}^{+-} ({\bf k}) = \sum_{L<L'} J_{\alpha}^{LL'} ({\bf k}) J_{\beta}^{L'L} ({\bf k}) 
\label{eq36} 
\end{equation}
and
\begin{equation}
J_{\alpha}^{+-} ({\bf k}) J_{\beta}^{-+} ({\bf k}) = \sum_{L>L'} J_{\alpha}^{LL'} ({\bf k}) J_{\beta}^{L'L} ({\bf k}).
\label{eq37} 
\end{equation}
	For $\alpha = \beta$, the result is of the form
\begin{equation}
\big| J_{\alpha}^{-+} ({\bf k}) \big|^2 = 2 \big(e v_{{\rm F}\alpha} \big)^2 \bigg[1 - \frac{K_\alpha^2}{K^2+M^2}\bigg],
\label{eq38} 
\end{equation}
while for $\alpha \neq \beta$, the coherence factors are odd functions of $K_\alpha$ and $K_\beta$; for example, 
\begin{equation}
J_{\alpha}^{-+} ({\bf k}) J_{\beta}^{+-} ({\bf k}) = - 2 v_{{\rm F}\alpha}v_{{\rm F}\beta} \frac{K_\alpha K_\beta}{K^2+M^2}.
\label{eq39} 
\end{equation}

For $n_L({\bf k})=f_L({\bf k})$, the resulting dynamical conductivity is given by
\begin{eqnarray}
&& \hspace{-8.5mm}
\sigma_{\alpha \alpha} ({\bf q},\omega) \approx 
\sigma_{\alpha \alpha} (\omega) = \sigma_{\alpha \alpha}^{\rm intra} (\omega) 
+ \sigma_{\alpha \alpha}^{\rm inter} (\omega),
\label{eq40}
\end{eqnarray}
where the intraband and interband contributions are given, respectively, by
\begin{eqnarray}
&& \hspace{-10mm}
\sigma_{\alpha \alpha}^{\rm intra} (\omega) = \frac{{\it i}e^2}{m} 
\frac{n_{\alpha \alpha}^{\rm intra}}{\omega +  i \Gamma_{1\alpha}},
\label{eq41}  \\
&& \hspace{-3mm}
n_{\alpha \alpha}^{\rm intra} = \frac{1}{V} {\sum_{L{\bf k}}}^* m [v_\alpha^L({\bf k})]^2
\bigg(- \frac{\partial f_L({\bf k})}{\partial \varepsilon_L ({\bf k})}\bigg)
\label{eq42}
\end{eqnarray}
and
\begin{eqnarray}
&& \hspace{-5mm} 
\sigma_{\alpha \alpha}^{\rm inter} ( \omega) =
\frac{2}{V}  {\sum_{{\bf k}}}^*
\frac{{\it i} \hbar |J^{-+}_{\alpha}({\bf k})|^2}{2\varepsilon({\bf k})}
\nonumber \\
&& \hspace{5mm}
\times \bigg[\frac{f_3({\bf k})-f_{1}({\bf k})}{\hbar \omega + 2\varepsilon({\bf k}) + {\it i}\hbar \Gamma_{2\alpha}}
+ \frac{f_3({\bf k})-f_{1}({\bf k})}{\hbar \omega - 2\varepsilon({\bf k}) + {\it i}\hbar \Gamma_{2\alpha}} \bigg].
\nonumber \\
\label{eq43}
\end{eqnarray}
    For simplicity,  we use here the approximation already used in the discussion of the dc conductivity in graphene in Sec.~II.
	In this approaximation, the intraband relaxation rate
is the same for all bands but can depend on the direction of electromagnetic field, 
i.e. $\Gamma^{LL}_\alpha ({\bf k}) \approx \Gamma_{1\alpha} = 1/\tau_\alpha$.
	Similarly, for the interband relaxation rates we assume that 
$\Gamma^{LL'}_\alpha ({\bf k}) \approx \Gamma_{2\alpha}$ ($L \neq L'$).
    The expressions (\ref{eq41}) and (\ref{eq43}) are an obvious generalization of the well-known
results characterizing conduction electrons in heavily doped graphene to the 3D case with the finite Dirac mass.

The square of the current vertices in Eqs.~(\ref{eq42}) and (\ref{eq43}) is proportional to $v_{{\rm F}\alpha}^2$.
    Moreover, the dispersions (\ref{eq21}) possess the spherical symmetry in the ${\bf K}$ space.
    Therefore, if we neglect the anisotropy in the relaxation rates, we obtain 
\begin{eqnarray}
&& \hspace{-5mm} 
\sigma_{\alpha \alpha} ( \omega) = \frac{v_{{\rm F}\alpha}^2}{\bar v_{\rm F}^2} \sigma_{\alpha \alpha}^{\rm iso} ( \omega).
\label{eq44}
\end{eqnarray}
    Here  $\bar v_{\rm F}$ is a useful abbreviation, $\bar v_{\rm F}^3= v_{{\rm F}x}v_{{\rm F}y}v_{{\rm F}z}$, and
    $\sigma_{\alpha \alpha}^{\rm iso} ( \omega)$ is the isotropic dynamical conductivity of the related isotropic problem in which  $v_{\rm F} = \bar v_{\rm F}$ is the isotropic Fermi velocity.
    On one hand, this simple relation is very useful in analyzing experimental data.
    If the ratios $\sigma_{\alpha \alpha} ( \omega)/\sigma_{\beta \beta} ( \omega)$ are nearly independent of frequency, then the relaxation rates  $\Gamma_{1\alpha}$ and  $\Gamma_{2\alpha}$ are nearly isotropic.
    The same conclusion holds for the temperature dependence of the dc conductivity.
From the theoretical standpoint, on the other hand, $\sigma_{\alpha \alpha}^{\rm iso} ( \omega)$ is interesting because the angular part of integration in the ${\bf K}$ space is trivial.
    
\section{Comparison with experiments}

\begin{figure}
   \centerline{\includegraphics[width=17pc]{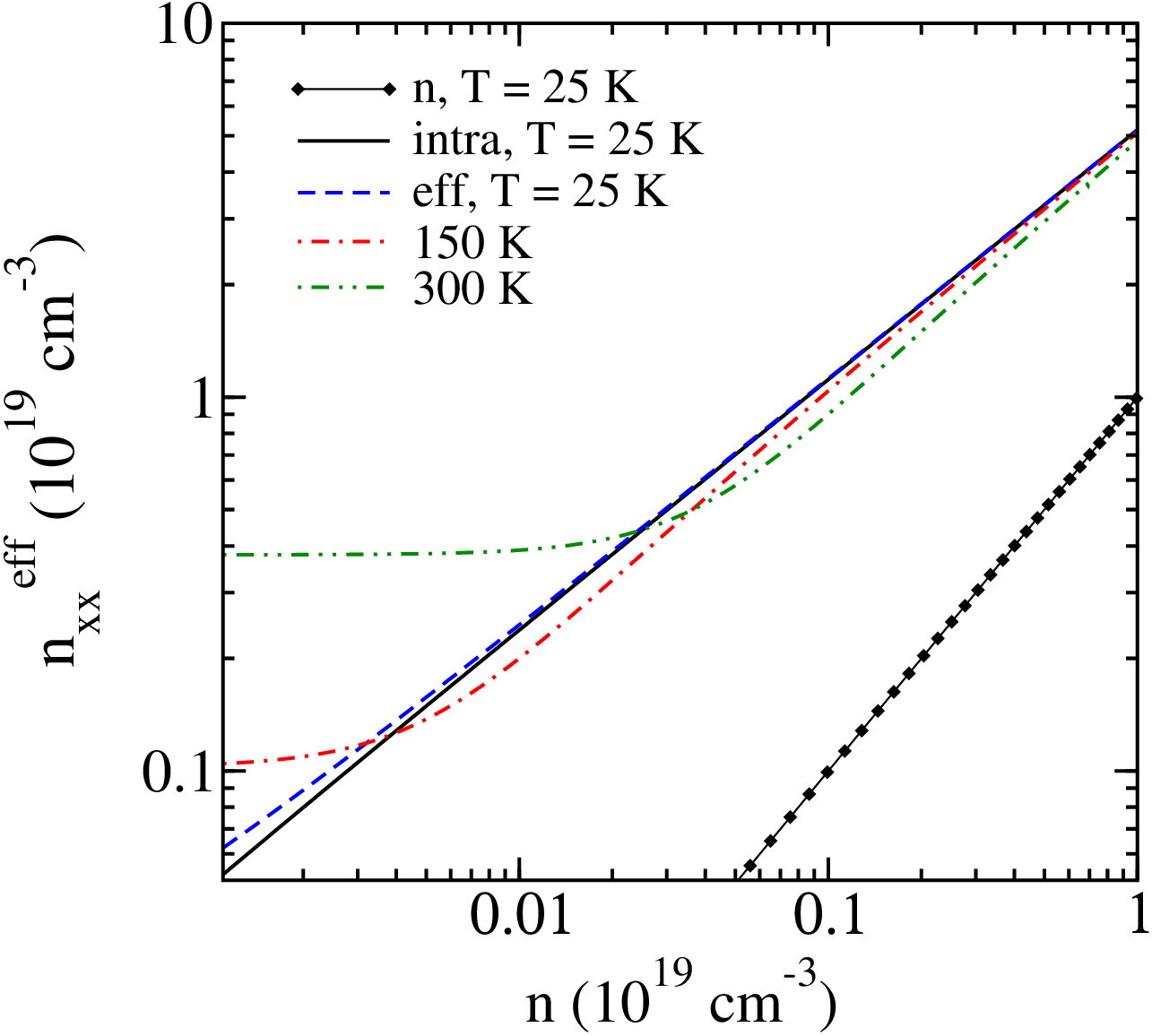}}
   \caption{(Color online)  The doping dependence of $n^{\rm eff}_{\alpha \alpha}$ in the isotropic ordinary massless 3D Dirac model for $v_{\rm F} = 4 \times 10^5$ ${\rm m}/{\rm s}$ and $\hbar \Gamma_1 = \hbar \Gamma_2 = 5$ meV at $T = 25$, 150, and 300 K. 
   The solid line is the intraband contribution calculated at $T= 25$ K.
}
  \end{figure} 	
  
Let us first consider the isotropic massless case with the Fermi velocity $v_{\rm F} = 4 \times 10^5$ ${\rm m}/{\rm s}$.
    Figure 3 shows the doping dependence of $n^{\rm eff}_{\alpha \alpha}$ [defined by Eq.~(\ref{eq2})] in the lightly doped 
region for typical values of model parameters at temperatures up to room temperature.
    The solid line is the low-temperature intraband contribution $n^{\rm intra}_{\alpha \alpha}$, while the diamonds represent $n$.
From this figure, one can see clearly that
    for the doping level $n > n_{\rm c}(300 \, {\rm K}) \approx 2 \times 10^{17}$ cm$^{-3}$,  the thermally activated contributions to $n^{\rm intra}_{\alpha \alpha}$ can be safely neglected.
    In this doping range the temperature dependence of the dc conductivity originates from the temperature dependence 
of the intraband relaxation rate.
    The integrated intraband conductivity spectral weight is also nearly temperature independent.
    
Let us now present
the main qualitative features of the real part of the dynamical conductivity in a typical anisotropic case with the $z$-axis anisotropy described by $\bar v_{\rm F} = 4 \times 10^5$ ${\rm m}/{\rm s}$ and
$v_{{\rm F}x} = 5 \times 10^5$ ${\rm m}/{\rm s}$, for two doping levels close to that found in two TlBiSSe samples (samples $S_1$ and $S_3$ in Ref.~\cite{LeMardele23}).
With little loss of generality, we restrict the calculation to the case where
both the intraband and interband relaxation rates are isotropic.

\begin{figure}
   \centerline{\includegraphics[width=17pc]{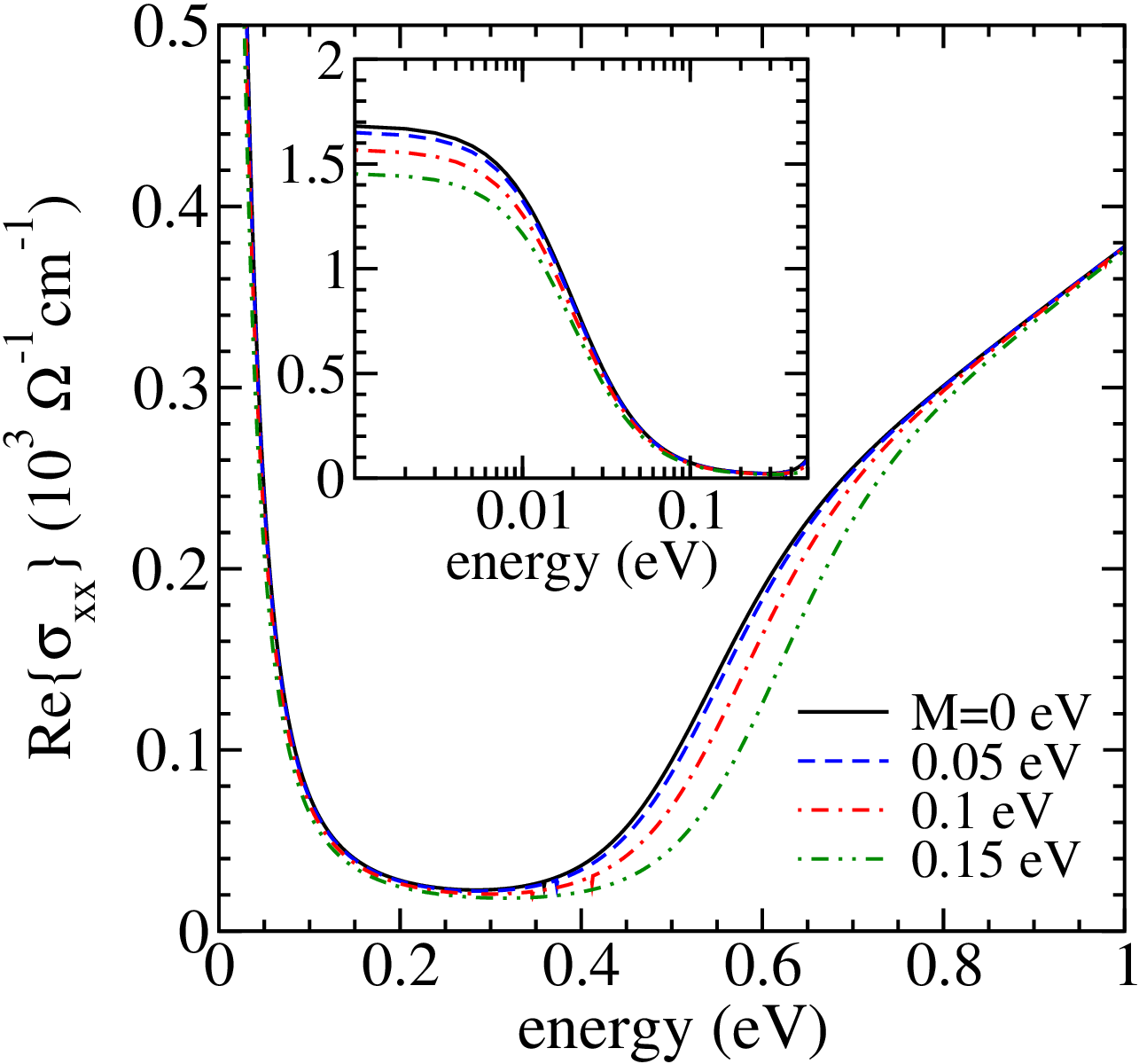}}
   \caption{(Color online) The real part of isotropic $\sigma_{xx}^{\rm iso} ( \omega)$ multiplied by $25/16$, Eq.~(\ref{eq44}), calculated for different values of the Dirac mass parameter $M$, for $\bar v_{\rm F} = 4 \times 10^5$ ${\rm m}/{\rm s}$, $\mu (300\,{\rm K}) = 0.26$ eV, $\hbar \Gamma_1 = \hbar \Gamma_2 = 20$ meV, and $T = 300$ K. 
   Inset: ${\rm Re} \{ \sigma_{xx} ( \omega) \}$ shown
   on a logaritmic scale.
}
  \end{figure} 	
    
The solid line in Fig.~4 shows the room-temperature in-plane dynamical conductivity ${\rm Re} \{ \sigma_{xx} ( \omega) \}$ for $M=0$, which agrees reasonably well with the spectrum measured on sample $S_1$.
    By fitting the intraband part of the spectrum, we obtain $\hbar \Gamma_1 \approx 20$ meV and 
$n^{\rm eff}_{xx} \approx 18.4 \times 10^{19}$ cm$^{-3}$.
    For given values of $\bar v_{\rm F}$ and $v_{{\rm F}x}$, we also obtain the nominal concentration of charge carriers
\begin{eqnarray}
&& \hspace{-5mm} 
n(T) = n (T=0) = \frac{K_{\rm F}^3}{3 \pi^2 \hbar^3 \bar v_{\rm F}^3} \approx 3.57 \times 10^{19}\, {\rm cm}^{-3}
\label{eq45}
\end{eqnarray}
and the chemical potential $\mu (300\,{\rm K}) = 0.26$ eV at $T=300$ K (heavily doped case).
    In the present case, the $z$-axis dynamical conductivity is about $25 \%$ of the in-plane conductivity ($v_{{\rm F}z} \approx 2.56 \times 10^5$ ${\rm m}/{\rm s}$).

The interband contribution to the dc conductivity is negligible even for a relatively large damping energy $\hbar \Gamma_2$
($\hbar \Gamma_2 = 20$ meV in the figure).
    The corresponding contribution to the dynamical conductivity increases with increasing $\hbar \Gamma_2$.
    It becomes dominant for $\hbar \omega > K_{\rm F}$.

It is well-known \cite{Kupcic16} that the other two Kubo formulas, when applied to gapless multiband electronic systems, give different behaviors of the in-gap interband conductivity. 
    The charge-charge Kubo formula underestimates and the current-current Kubo formula overestimates the low-frequency value of 
${\rm Re} \{ \sigma_{\alpha \alpha}^{\rm inter} ( \omega) \}$.
    As discussed in Appendix D, we use here the current-dipole Kubo formula because it is consistent with both the transverse conductivity sum rule and the related effective mass theorem.

    \begin{figure}
   \centerline{\includegraphics[width=17pc]{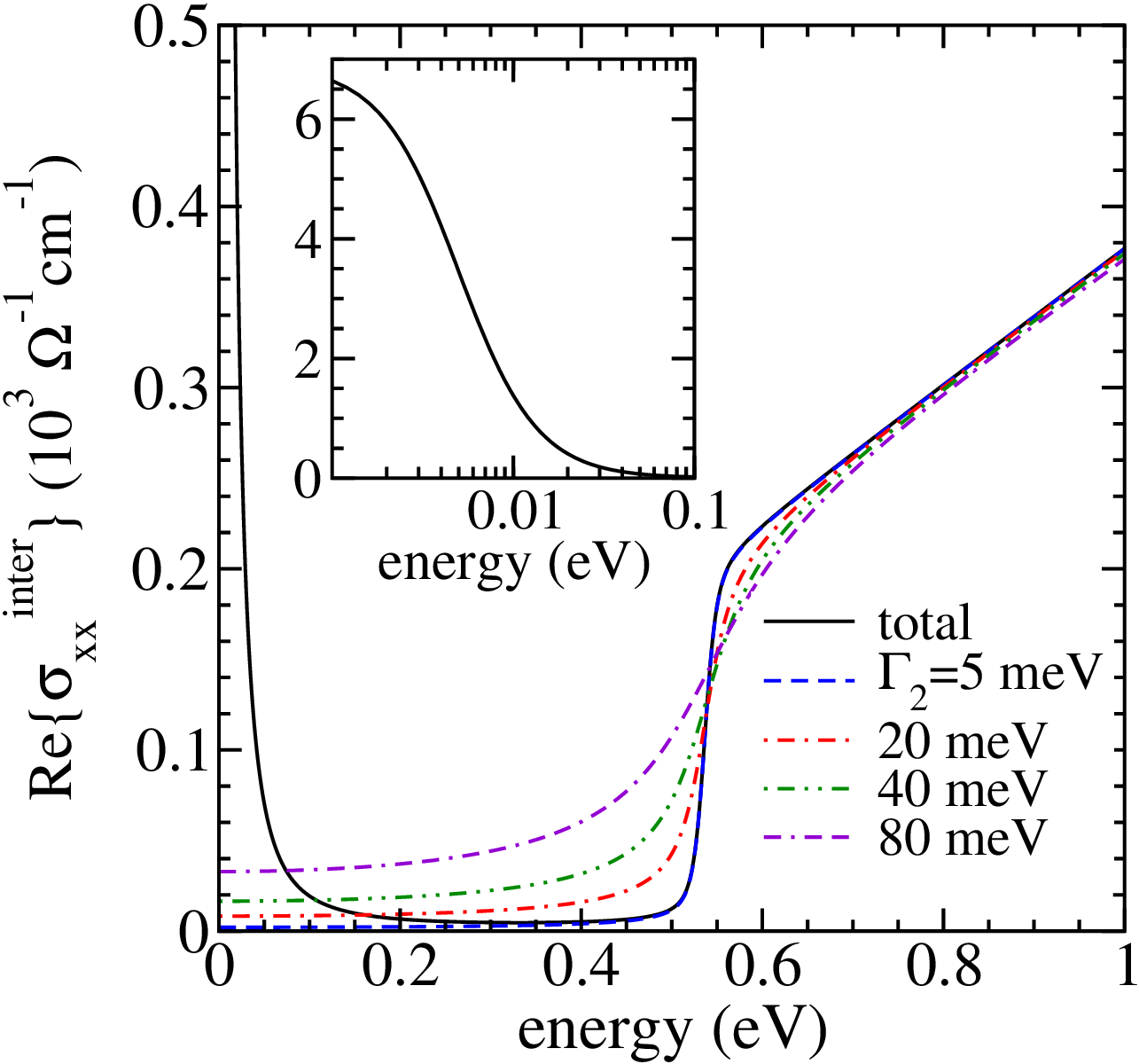}}
   \caption{(Color online) The real part of $\sigma_{xx}^{\rm inter} ( \omega)$ calculated for $\hbar \Gamma_1 = 5$ meV and $T = 25$ K, for different values of $\hbar \Gamma_2$.
   The other parameters are the same as in Fig.~4.
   Inset: ${\rm Re} \{ \sigma_{xx} ( \omega) \}$ shown on a logaritmic scale.
}
  \end{figure} 	

The comparison with the interband part of the spectrum measured in the energy range $0.2-0.5$ eV shows that in this energy range the 
interband contribution (\ref{eq43}) accounts only for $50 \%$ of the observed intensity.
    This is not surprising because the {\it ab initio} calculations \cite{LeMardele23}
show that interband optical excitions that involve the states from the
rest of the Brillouin zone start already at the energy close to 0.5 eV.

Figure 4 also illustrates how ${\rm Re} \{\sigma_{xx} ( \omega) \}$ changes with changing the Dirac mass parameter $M$, 
for $M/K_{\rm F}$ not too large.
    Since the concentration (\ref{eq45}) depends only on the bare Fermi energy $K_{\rm F}$, the zero-temperature interband threshold energy shifts with $M$ as $2\sqrt{K_{\rm F}^2+M^2}$.
    
Figure 5 illustrates the dependence of low-temperature ${\rm Re} \{\sigma_{xx} ( \omega) \}$ on $\hbar \Gamma_2$ for $M=0$ and $\mu (300 \, {\rm K}) = 0.26$ meV. 
    The intraband relaxation rate $\hbar \Gamma_1 \approx 5$ meV is estimated from measured resistivity data. \cite{Novak15}
    The figure shows that for $n \approx 3.57 \times 10^{19}$ cm$^{-3}$, even for $\hbar \Gamma_2 = 40$ meV, the interband contribution to $n^{\rm eff}_{xx}$ is below $0.5 \%$.
  
In experimental studies, it is common to represent the integrated conductivity spectral weight by the spectral function $N_{\alpha \alpha}(\omega)$ defined by
\begin{equation}
N_{\alpha \alpha}(\omega) = \sum_i N_{\alpha \alpha}^{i}(\omega) =
\frac{8}{\Omega_0^2} \int_0^\omega {\rm d} \omega' {\rm Re} \{ \sigma_{\alpha \alpha} ( \omega') \}. 
\label{eq46}
\end{equation}
    Here, $\Omega_0 = \sqrt{ 4 \pi e^2/mV_0}$ is an auxiliary plasma frequency, $V_0$ is the related auxiliary primitive cell volume, and the index $i =$ intra, inter.
    For $V_0 = 100$ \AA $^3$, $N_{\alpha \alpha}^{\rm intra}(\omega \gg \Gamma_1)$ is equal to $n_{\alpha \alpha}^{\rm intra}$, given in units of $10^{19}$ cm$^{-3}$.
    Figure 6 illustrates the spectral function $N_{xx}(\omega)$ for the spectra shown in Fig.~5.

      \begin{figure}
   \centerline{\includegraphics[width=17pc]{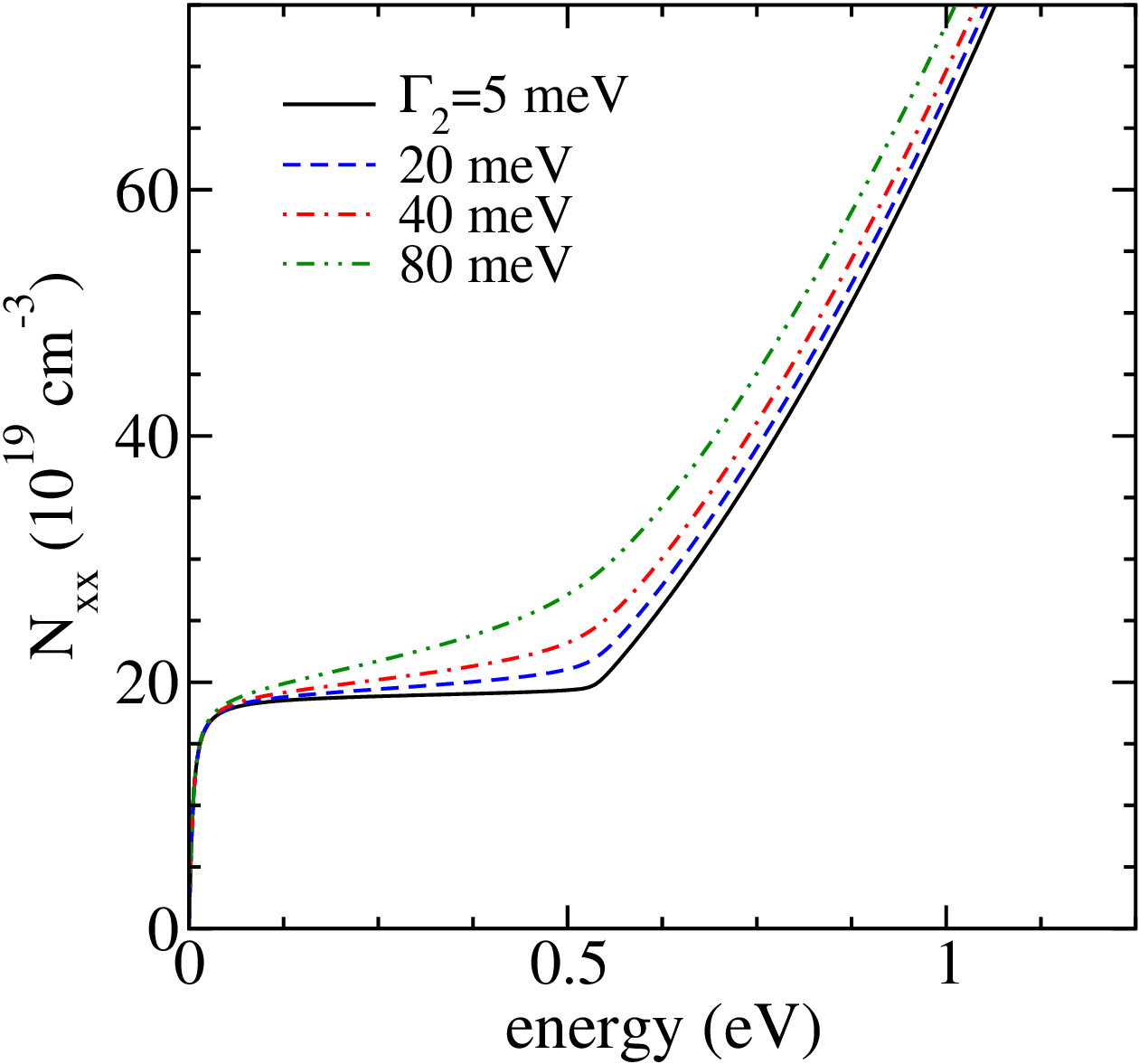}}
   \caption{(Color online) The dependence of the spectral function $N_{xx}(\omega)$ on $\hbar \Gamma_2$ for the spectra shown in Fig.~5.
}
  \end{figure}

It is important to notice that in the Dirac cone approximation this spectral function is well-defined as long as the upper limit of integration over $\omega'$ is below the cut-off energy used in the restricted sum $\sum_{\bf k}^*$ in Eqs.~(\ref{eq42}) and (\ref{eq43}).
    When $\hbar \omega$ is above this cut-off energy, then $N_{xx}(\omega)$ saturates to $n_{\alpha \alpha}^{\rm total}$ given by (\ref{eqD7}).

In the present context, the most important fact about
sample $S_3$ from Ref.~\cite{LeMardele23} is that the in-plane dc conductivity is one order of magnitude smaller than the dc conductivity measured previously in a similar sample.
    This is consistent with the conclusions of previous theoretical studies of lightly doped Dirac systems in the dirty regime.
\cite{Carbotte10,Tabert16,Kupcic16}
    In these studies, it is shown that for the doping level $n < n_{\rm c}(300 \, {\rm K})$ the damping effects in $\sigma_{\alpha \alpha} ( \omega)$ have to be treated beyond the $n_L({\bf k})=f_L({\bf k})$ approximation.
    In this case, the momentum distribution function is given by the general expression \cite{Mahan90,Kupcic16}
\begin{eqnarray}
&& \hspace{-10mm}
n_L({\bf k}) =  \int_{-\infty}^\infty \frac{d \varepsilon}{2 \pi} \, {\cal A}_L ({\bf k}, \varepsilon) f(\varepsilon),
\label{eq47}
\end{eqnarray}
where 
\begin{eqnarray}
&& \hspace{-5mm}
{\cal A}_L ({\bf k}, \varepsilon) \approx  \frac{2 \hbar \Sigma^i_L}{
[\varepsilon - \varepsilon_L ({\bf k}) ]^2 + [\hbar \Sigma^i_L ]^2}
\label{eq48}
\end{eqnarray}
is the single-electron spectral function in question and $\hbar \Sigma^i_L \approx \hbar \Sigma^i$ is the corresponding 
single-electron damping energy.   
    The dependence of the dc conductivity of dirty lightly doped Dirac semimetals on $\hbar \Sigma^i$ is expected to be similar 
to that found in dirty lightly doped graphene (Fig.~7 in Ref.~\cite{Kupcic16}).
    A detailed discussion of this subject will be given in a future presentaion \cite{KupcicUP}.

\begin{figure}
   \centerline{\includegraphics[width=17pc]{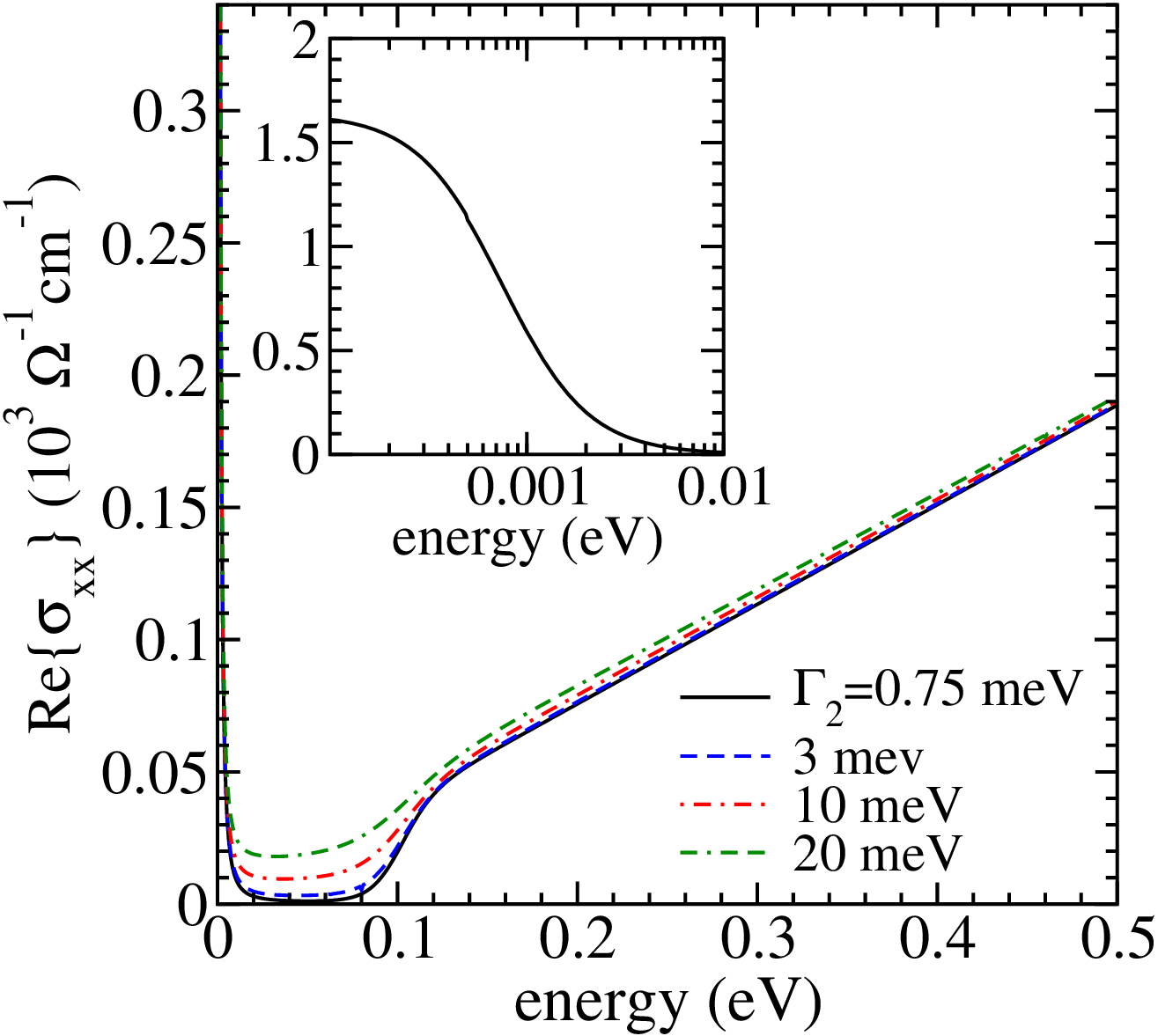}}
   \caption{(Color online) The dependence of ${\rm Re} \{\sigma_{xx} ( \omega) \}$ on $\hbar \Gamma_2$ for $M = 0$, $\mu (50 \, {\rm K}) = 50$ meV, $\hbar \Gamma_1 = 0.75$ meV, and $T=50$ K.
   Inset: ${\rm Re} \{ \sigma_{xx} ( \omega) \}$ shown on a logaritmic scale.
}
  \end{figure}     
  
\begin{figure}
   \centerline{\includegraphics[width=17pc]{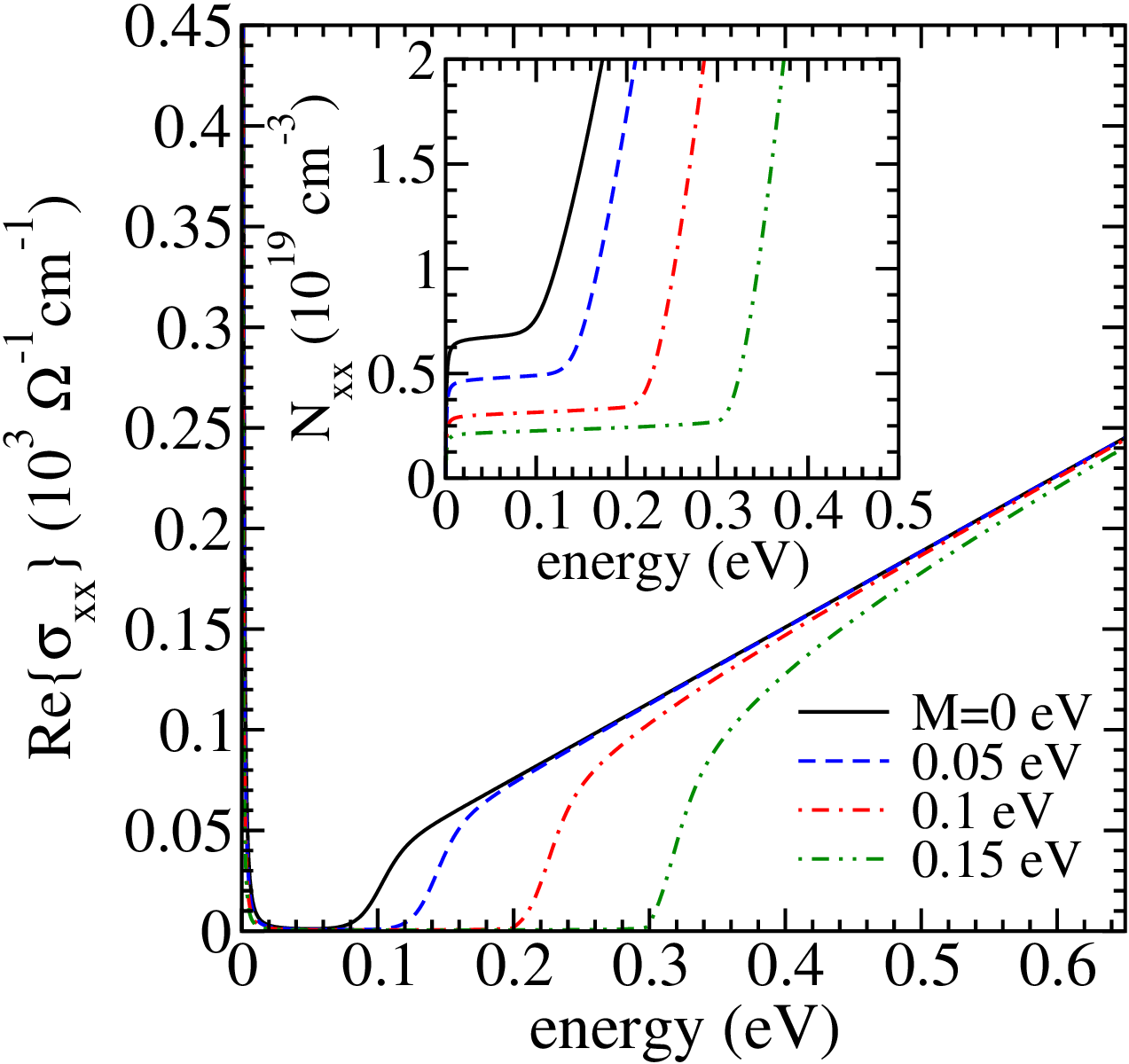}}
   \caption{(Color online) The dependence of ${\rm Re} \{\sigma_{xx} ( \omega) \}$ on $M$ for $\mu (50 \, {\rm K}) = 50$ meV, $\hbar \Gamma_1 = \hbar \Gamma_2 = 0.75$ meV, and $T=50$ K.
   Inset: The spectral function $N_{xx}(\omega)$ for the same spectra.
}
  \end{figure}    

In order to show the effects of $\hbar \Gamma_2$ and $M$ on ${\rm Re} \{ \sigma_{xx} ( \omega) \}$ in the case in which these two energy scales are comparable to the interband threshold energy, we take the doping level
$n = 2.5 \times 10^{17} \, {\rm cm}^{-3} \approx n_{\rm c}(300 \, {\rm K})$  as an example [$\mu (50 \, {\rm K}) = 50$ meV in this case].
    Figures 7 and 8 illustrate the real part of $\sigma_{xx} ( \omega)$ for different values of $\hbar \Gamma_2$ and $M$, respectively.
    The inset of Fig.~8 shows the spectral function $N_{xx}(\omega)$ for the same values of $M$.
    Notice that the effective number $n_{xx}^{\rm eff} = 6.7 \times 10^{18} \, {\rm cm}^{-3}$ for $M=0$ is by a factor of 27 larger than $n$.
    It decreases with increasing $M$ [$n_{xx}^{\rm eff} \approx  N_{xx}(\omega \approx K_{\rm F})$ in the inset of figure].

\section{Electron mobility}
Finally, let us briefly discuss two expressions for the electron mobility from Sec.~II.
    For simplicity, we consider a heavily doped case, where interband contributions to $n_{\alpha \alpha}^{\rm eff}$ are negligible.
    It is convenient to define an averaged reciprocal effective mass $\langle \gamma_{\alpha\alpha}^{LL} ({\bf k}) \rangle/m$ in the following way 
\begin{eqnarray}
&& \hspace{-5mm} \frac{n^{\rm intra}_{\alpha \alpha}}{m} = \frac{1}{V} {\sum_{L{\bf k}}} 
\frac{\gamma_{\alpha\alpha}^{LL} ({\bf k})}{m} f_L({\bf k}) =\frac{ \langle \gamma_{\alpha\alpha}^{LL} ({\bf k}) \rangle}{m} n.
\label{eq49}
\end{eqnarray}
    It is also useful to recall that the cyclotron mass in the electron doped anisotropic ordinary 3D Dirac model is given by
\begin{eqnarray}
&& \hspace{-5mm} m^*_{\alpha} (\varepsilon, k_z) = m^*_{\alpha} (\varepsilon) = \frac{v_{{\rm F}\alpha}}{\bar v_{\rm F}^3} \varepsilon.
\label{eq50}
\end{eqnarray}
    At zero temperature, the relation between these two expressions is the following
\begin{eqnarray}
&& \hspace{-5mm} \frac{m}{\langle \gamma_{\alpha\alpha}^{LL} ({\bf k}) \rangle_0} =  \frac{\bar v_{\rm F}^3}{v_{{\rm F}\alpha}^3}m^*_{\alpha} (\varepsilon_{\rm F}).
\label{eq51}
\end{eqnarray}
    This means that $\mu^{\rm exp}$ and $\mu^{\rm th}$ from Sec.~II represent the mobility of conduction electrons with mass $m/\langle \gamma_{\alpha\alpha}^{LL} ({\bf k}) \rangle$ and $m$, respectively  [in the isotropic case at zero temperature the former mass is equal to $m^*_{\alpha} (\varepsilon_{\rm F})$].
    The corresponding concentrations are $n$ and $n_{\alpha \alpha}^{\rm intra}$.

Therefore, in heavily doped 3D Dirac semimetals at low temperatures there is no difference between these two representations of conduction electrons.
    However, it is not obvious how to interpret the interband contribution to the dc conductivity in lightly doped samples [Eqs.~(\ref{eq2}) and (\ref{eq3})] in terms of different effective masses, or how to understand different plasma oscillations in the transverse conductivity sum rule from Appendix D [Eqs.~(\ref{eqD2}) and (\ref{eqD4})], also in terms of different effective masses.
    
\section{Conclusion}
In this paper, we have rederived the gauge-invariant tight-binding minimal substitution in a general noninteracting multiband case in which the electrons are shown in the representation of delocalized molecular orbitals.
    The exact expressions for the electron-photon coupling functions (current vertices and bare diamagnetic vertices) are used then 
to determine the elements of the current-dipole dynamical conductivity tensor. 
    This conductivity formula is known to be consistent with the charge continuity equation.
    Here, it is shown that it is consistent with the effective mass theorem, as well.
 
The results are applied to lightly doped and heavily doped anisotropic 3D Dirac semimetals.
    The model parameters used in numerical calculations are obtained by fitting the resistivity and reflectivity data measured 
on two TlBiSSe samples.
The key to quantitative understanding of 
measured data is to make clear distinction between the effective number 
of charge carriers $n^{\rm eff}_{\alpha\alpha}$ and their nominal concentration $n$.
    Although these two numbers represent essentially the same physical quantity in simple electronic systems with parabolic 
dispersion, here the ratio $n^{\rm eff}_{xx}/n$ is found to increases from $\approx 4$ to $\approx 25$ when the doping level is changed from $n \approx 3.6 \times 10^{19} \, {\rm cm}^{-3}$ to $n \approx 2.5 \times 10^{17} \, {\rm cm}^{-3}$.
    The momentum distribution function is found to play an important role in explaining differences between heavily doped 
and lightly doped samples.

\section*{Acknowledgments}
This research was supported by the University of Zagreb Grant No. 20286570

\appendix	

\section{Tight-binding minimal substitution}
The gauge-invariant tight-binding minimal substitution
is valid under quite general conditions. \cite{Schrieffer64,Kupcic07,Kupcic14}
    It is applicable to a rich variety of problems.
    It represents a widely used tool for investigating 
electrodynamic properties of valence electrons described by different types of noninteracting multiband electronic models.
    In all such cases, an appropriate starting point is
the bare Hamiltonian shown in the representation of orthogonal atomic orbitals (Wannier functions)
\begin{eqnarray}
&& \hspace{-5mm}
H_0 =  \sum_{ll'}\sum_{n n'}\sum_{\sigma  \sigma'} c^\dagger_{l n  \sigma} \langle l n  \sigma | H_0 | l' n' \sigma' \rangle c_{l' n' \sigma'}.
\label{eqA1}
\end{eqnarray}
	Here $c^\dagger_{ln  \sigma}$ is the electron creation operator in the atomic orbital labeled by the orbital index 
$l$ placed at the lattice site ${\bf R}_n + {\bf r}_l$.
	The matrix elements $\langle l n  \sigma | H_0 | l' n' \sigma' \rangle$ stand for all relevant site energies and bond energies.

In order to examine how electrons in Eq.~(\ref{eqA1}) respond to applied electromagnetic fields,
we use the  substitution $c^\dagger_{l n \sigma} \rightarrow \widetilde c^\dagger_{l n \sigma}$ in $H_0$, where
\begin{eqnarray}
&& \hspace{-10mm}
\widetilde c^\dagger_{l n \sigma} =  e^{i(e/\hbar c) {\bf A} ({\bf R}_n+{\bf r}_l) \cdot ({\bf R}_n+{\bf r}_l)} 
c^\dagger_{l n \sigma} .
\label{eqA2}
\end{eqnarray}
	The resulting total Hamiltonian 
\begin{eqnarray}
&& \hspace{-5mm}
\widetilde H_0 =  \sum_{ll'}\sum_{n n'}\sum_{\sigma  \sigma'} \widetilde c^\dagger_{l n  \sigma} \langle l n  \sigma | H_0 | l' n' \sigma' \rangle \widetilde c_{l' n' \sigma'}
\label{eqA3}
\end{eqnarray}
is the sum of the bare Hamiltonian $H_0$ and the coupling Hamiltonian $H^{\rm ext}$.
	
To obtain the alternative form of $\widetilde H_0$
it is useful first to show $H_0$ in the representation of delocalized atomic orbitals 
\begin{eqnarray}
&& \hspace{-10mm}
c^\dagger_{l {\bf k} \sigma} = \frac{1}{\sqrt{N}} \sum_{n} e^{i {\bf k} \cdot {\bf R}_n} c^\dagger_{l n \sigma}.
\label{eqA4}
\end{eqnarray}
	The result is 
\begin{eqnarray}
&& \hspace{-5mm}
H_0 =  \sum_{ll'}\sum_{{\bf k} \sigma\sigma'}  H_0^{l\sigma, l'\sigma'} ({\bf k}) 
c^\dagger_{ l{\bf k} \sigma} c_{l'{\bf k} \sigma'},
\label{eqA5}
\end{eqnarray}
with $H_0^{l\sigma, l'\sigma'} ({\bf k})$ simply related to $\langle l n  \sigma | H_0 | l' n' \sigma' \rangle$.
	This expression can be rewritten in terms of the momentum operator  
$\hat {\bf p}_{n}  = (\hbar/i) \partial / \partial {\bf R}_n$, in the following way
\begin{eqnarray}
&& \hspace{-5mm}
H_0 =   \sum_{ll'}\sum_{n n'}\sum_{\sigma \sigma'} c^\dagger_{ln  \sigma} \delta_{{\bf R}_n,{\bf R}_{n'}} 
H_0^{l\sigma,l'\sigma'}(\hat {\bf p}_{n'})c_{l'n' \sigma'}.
\label{eqA6}
\end{eqnarray}
    The total Hamiltonian $\widetilde H_0$ is given now by the latter expression in which the matrix elements 
$H_0^{l\sigma,l'\sigma'}(\hat {\bf p}_{n'})$ are replaced by 
$H_0^{l\sigma,l'\sigma'}(\hat {\bf p}_{n'} - e/c {\bf A} ({\bf R}_{n}))$.
	The result is 
\begin{equation}
\widetilde H_0 =   \sum_{ll'}\sum_{n n'}\sum_{\sigma \sigma'} c^\dagger_{ln  \sigma} \delta_{{\bf R}_n,{\bf R}_{n'}} 
H_0^{l\sigma,l'\sigma'}(\hat {\bf p}_{n'}- \frac{e}{c} {\bf A} ({\bf R}_{n}))c_{l'n' \sigma'}.
\label{eqA7}
\end{equation}

The expressions (\ref{eqA3}) and (\ref{eqA7}) lead to the same expression for $H^{\rm ext}$, for the current vertex functions, and for the dynamical conductivity tensor, as long as the states $c^\dagger_{l {\bf k} \sigma} | 0 \rangle$ in Eq.~(\ref{eqA4}) represent delocalized atomic orbitals.

The 2D Dirac model is an example of such an exacly solvable multiband problem in which
the expressions (\ref{eqA3}) and (\ref{eqA7}) lead to the same form of $H^{\rm ext}$.
    This is a direct consequence of
the fact that all elements in (\ref{eqA1}) are simple functions of the first-neighbor bond energy $t$, two site energies, $E_A$ and $E_B$, and two second-neighbor bond energies, $t'_A$ and $t'_B$.

\section{Current vertices in the ordinary 3D Dirac model}
To obtain the coupling Hamiltonian between the conduction electrons and external electromagnetic fields in the ordinary 3D Dirac model, we perform the Taylor expansion of Eq.~(\ref{eq27}) in the main text,
\begin{eqnarray}
&& \hspace{-5mm}
\widetilde H_0 =   \sum_{ll'}\sum_{n n'} c^\dagger_{ln } \delta_{{\bf R}_n,{\bf R}_{n'}} 
H_0^{ll'}(\hat {\bf p}_{n'}- \frac{e}{c} {\bf A} ({\bf R}_{n}))c_{l'n'},
\label{eqB1}
\end{eqnarray}
to the second order in the vector potential.
    This expression for $\widetilde H_0$ is obtained from (\ref{eqA7}) by omitting spin indices and by replacing local atomic 
orbitals $c^\dagger_{ln \sigma} | 0 \rangle$ by local molecular orbitals $c^\dagger_{ln } | 0 \rangle$.
	The result is
\begin{eqnarray}
&& \hspace{-5mm}
\widetilde H_0 = H_0 + H^{\rm ext},
\label{eqB2}
\end{eqnarray}
where 
\begin{eqnarray}
&& \hspace{-10mm}
H^{\rm ext} = H^{\rm ext}_1 + H^{\rm ext}_2 + \ldots
\nonumber \\   
&& \hspace{-2mm}
= \sum_{ll'}\sum_{{\bf k}{\bf q}}  \delta H_0^{ll'} ({\bf k},{\bf q}) c_{l{\bf k} + {\bf q}}^\dagger c_{l'{\bf k}}  + \ldots,
\label{eqB3}
\end{eqnarray}
with
\begin{eqnarray}
&& \hspace{-10mm}
\delta  H_0^{ll'} ({\bf k},{\bf q})  \approx  - \frac{e}{c}
\sum_{\alpha} \frac{\partial  H_0^{ll'} ({\bf k})}{\partial p_{\alpha}} A_{\alpha} ({\bf q}) 
\nonumber \\   
&& \hspace{7mm}
+ \frac{e^2}{2c^2} \sum_{{\bf q}' \alpha \beta}
\frac{\partial^2 H_0^{ll'} ({\bf k})}{\partial p_{\alpha} \partial p_{\beta}}
A_{\alpha}  ({\bf q}-{\bf q}') A_{\beta}  ({\bf q}') 
\label{eqB4} 
\end{eqnarray}
and $p_\alpha = \hbar k_\alpha$ again.

We use now the transformation matrix elements $U_{\bf k}(l,L)$ from the main text,
between the delocalized molecular orbitals $|l{\bf k}\rangle$  and the Bloch states $|L{\bf k}\rangle$,
\begin{eqnarray}
&& \hspace{-10mm}
|l{\bf k} \rangle  = \sum_L  U_{\bf k}(l,L) |L{\bf k} \rangle , 
\label{eqB5}
\end{eqnarray}
to obtain
\begin{eqnarray}
&& \hspace{-5mm}
J_{\alpha}^{LL'} ({\bf k},{\bf k}_+) \approx J_{\alpha}^{LL'} ({\bf k}) = 
\sum_{ll'} U_{\bf k}^T(L,l)j_{\alpha}^{ll'} ({\bf k})  U_{\bf k}^*(l',L').
\nonumber \\
\label{eqB6}
\end{eqnarray}

As mentioned in the main text, in the ordinary 3D Dirac model both dispersive corrections  in Eq.~(\ref{eq13}) are set to zero 
[$\varepsilon_0({\bf k}) = \Delta M  ({\bf k}) =0$].
    In this case, the bare current vertex functions are given by
\begin{eqnarray}
&& \hspace{-10mm}
 j_{\alpha}^{ll'} ({\bf k}) = e  \frac{\partial  H_0^{ll'} ({\bf k})}{\partial p_{\alpha}}, 
\label{eqB7} 
\end{eqnarray}
while the auxiliary phase $\phi ({\bf k})$ satisfies the relation
\begin{eqnarray}
&& \hspace{-5mm}	
\tan \phi ({\bf k}) = \frac{K}{M}.
\label{eqB8}
\end{eqnarray}
    A straightforward calculation gives the following expressions for the intraband and interband current 
vertex functions $J_{\alpha}^{LL'} ({\bf k})$:
\begin{equation}
\bigg(  J_{x}^{LL'} ({\bf k}) \bigg) 
=  \frac{e v_{{\rm F}x}}{K}\left( \begin{array}{cccc} 
K_x S & 0 & - K_x C & -K_{x+} \\
0 & K_x S  & K_{x-} & - K_x C \\
- K_x C & K_{x+} & -K_x S & 0 \\
-K_{x-} & - K_x C & 0 & -K_x S
\end{array} \right),
\label{eqB9}
\end{equation}
\begin{equation}
\bigg(  J_{y}^{LL'} ({\bf k}) \bigg) 
=  \frac{e v_{{\rm F}y}}{K}\left( \begin{array}{cccc} 
K_y S & 0 & K_{y-} & iK_{x} \\
0 & K_y S  & iK_{x} & K_{y+} \\
K_{y+}  & -iK_{x} & -K_y S & 0 \\
-iK_{x} & K_{y-} & 0 & -K_y S
\end{array} \right),
\label{eqB10}
\end{equation}
and
\begin{equation}
\bigg(  J_{z}^{LL'} ({\bf k}) \bigg) 
=  \frac{e v_{{\rm F}z}}{K}\left( \begin{array}{cccc} 
K_z S & 0 & K_{z+} & K_{x} \\
0 & K_z S  & -K_{x} & K_{z-} \\
K_{z-}  & -K_{x} & -K_z S & 0 \\
K_{x} & K_{z+} & 0 & -K_z S
\end{array} \right).
\label{eqB11}
\end{equation}
	In these expressions, we use the abbreviations 
$S = \sin \phi ({\bf k})$, $C = \cos \phi ({\bf k})$, $K_{x\pm} =K_z \pm i K_y$, 
$K_{y\pm} =\pm i K_z - K_y \cos \phi ({\bf k})$, and $K_{z\pm} =\pm i K_y - K_z \cos \phi ({\bf k})$.

The intraband current vertices are directly related with the corresponding electron group velocities, 
\begin{equation}
J_{\alpha}^{LL} ({\bf k}) = s_L e v_{{\rm F}\alpha} \sin \phi ({\bf k}) \frac{K_\alpha}{K} \equiv e v_{\alpha}^{L} ({\bf k}) .
\label{eqB12} 
\end{equation}
	Moreover, the interband current vertices between the bands that are degenerate 
in energy are equal to zero, 
\begin{equation}
J_{\alpha}^{12} ({\bf k}) = J_{\alpha}^{21} ({\bf k}) = J_{\alpha}^{34} ({\bf k}) = J_{\alpha}^{43} ({\bf k}) = 0.
\label{eqB13} 
\end{equation}

\section{Missing contributions to $\sigma_{\alpha \alpha} (\omega)$}
It is tempting to combine the procedure from Appendix B with other two forms of the bare Hamiltonian of the ordinary 3D Dirac model, Eqs.~(\ref{eq20}) and (\ref{eq17}), 
\begin{eqnarray}
&& \hspace{-10mm}
H_0 = \sum_{L{\bf k}} \varepsilon_L ({\bf k}) c^\dagger_{L {\bf k}} c_{L {\bf k}},
\nonumber \\
&& \hspace{-10mm}
H_0 =  \sum_{mm'{\bf k} }  H_0^{mm'} ({\bf k}) c^\dagger_{ m{\bf k}} c_{m'{\bf k}}.
\label{eqC1}
\end{eqnarray}
    The coupling Hamiltonian is given again by the expression (\ref{eq30}).
    However, the current vertices and the bare diamagnetic vertices have much simpler form.
    In the $\{ L {\bf k} \}$ representation, all interband contributions to $\sigma_{\alpha \alpha} (\omega)$ are missing, because
\begin{equation}
J_{\alpha}^{LL'} ({\bf k}) = \delta_{L,L'}  e \frac{\partial \varepsilon_L ({\bf k})}{\partial p_\alpha},
\label{eqC2} 
\end{equation}
in this case.
    In the $\{ m {\bf k} \}$ representation, some interband terms in $J_{\alpha}^{LL'} ({\bf k})$ are restored.
    In this case, the result is
\begin{eqnarray}
&& \hspace{-10mm} 
\left( J_{\alpha}^{LL'} ({\bf k})  \right) 
=  
\left( \begin{array}{cccc} 
 K_\alpha S {\rm I}_2 & -K_\alpha C{\rm I}_2 \\
- K_\alpha C{\rm I}_2 & - K_\alpha S {\rm I}_2
\end{array} \right).
\label{eqC3}
\end{eqnarray}

\section{Effective mass theorem and transverse conductivity sum rule}
One of the central questions regarding 
the transverse conductivity sum rule in multiband electronic systems is to establish relation between the integrated intraband and interband conductivity spectral weights and the effective mass theorem.
    The most important fact about
the conductivity sum rule and the effective mass theorem is that they are both
insensitive to details
in the intraband and interband relaxation rates.
    The well known $f$-sum rule \cite{Pines89,Ziman79,Kittel87} is a simple example of this general case.
    In this example, the number of bands is infinite, the electron effective mass is approximated by the ${\bf k}=0$ effective mass from the ${\bf k} \cdot {\bf p}$ perturbation theory, and the effective number $n_{\alpha \alpha}^{\rm total}$ in the total plasma frequency is equal to the concentration $n$.

In order to apply such an analysis on the 3D Dirac model, it is important to recall that in this model the bottom of lower bands is placed at negative infinity.
    This means that the sum over all occupied states in these bands must be evaluated in the hole picture.

In a general spinless multiband case, the bare diamagnetic vertex functions $\gamma^{LL}_{\alpha \alpha}({\bf k};2)$ are finite.
    They are known to satisfy the effective mass theorem \cite{Kupcic07}
\begin{eqnarray}
&& \hspace{-10mm}
\gamma^{LL}_{\alpha \alpha}({\bf k};2) = \gamma^{LL}_{\alpha \alpha}({\bf k}) 
+ \frac{m}{e^2} \sum_{L'(\neq L)} \frac{2 J_\alpha^{L L'}({\bf k}) J_\alpha^{L'L}({\bf k})}{
\varepsilon_{L'L}({\bf k},{\bf k})}.
\label{eqD1}
 \end{eqnarray} 
    The integrated total conductivity spectral weight is usually shown in the following way \cite{Kupcic03,Kupcic17B}
\begin{eqnarray}
&& \hspace{-5mm} 8 \int_0^\infty {\rm d} \omega {\rm Re} \{ \sigma_{\alpha \alpha} ( \omega) \} = \frac{4 \pi e^2 n_{\alpha \alpha}^{\rm total}}{m} = \Omega_{\rm total, \alpha}^2. 
\label{eqD2}
\end{eqnarray}
    After inserting Eq.~(\ref{eq33}) into this definition relation, we obtain
\begin{eqnarray}
&& \hspace{-5mm} n_{\alpha \alpha}^{\rm total} 
= \frac{1}{V} \sum_{L{\bf k}} \gamma_{\alpha \alpha}^{LL} ({\bf k};2) n_L({\bf k}).
\label{eqD3}
\end{eqnarray}
    $n_{\alpha \alpha}^{\rm total}$ represents the total effective number of charge carriers and $\Omega_{\rm total, \alpha}$ is the corresponding bare total plasma frequency.

The integrated intraband conductivity spectral weight is given by the textbook expression \cite{Pines89,Ziman79,Mahan90}
\begin{eqnarray}
&& \hspace{-5mm} 8 \int_0^\infty {\rm d} \omega {\rm Re} \{ \sigma_{\alpha \alpha}^{\rm intra} ( \omega) \} = \frac{4 \pi e^2 n_{\alpha \alpha}^{\rm intra}}{m} = \Omega_{\rm intra, \alpha}^2, 
\label{eqD4}
\end{eqnarray}
where $n_{\alpha \alpha}^{\rm intra}$ is given by Eq.~(\ref{eq42}) and $\Omega_{\rm intra, \alpha}$ is the bare intraband plasma frequency.

In the 3D Dirac model, the effective mass theorem, together with $\gamma^{LL}_{\alpha \alpha}({\bf k};2) = 0$, gives
\begin{eqnarray}
&& \hspace{-10mm}
\gamma^{LL}_{\alpha \alpha}({\bf k}) = s_L \frac{m}{e^2}  \frac{|J_\alpha^{+-}({\bf k})|^2}{\varepsilon_{13}({\bf k},{\bf k})} 
\nonumber \\
&& \hspace{2mm}
= s_L m \frac{v_{{\rm F}\alpha}^2}{\sqrt{K^2 + M^2}} \bigg[1 - \frac{K_\alpha^2}{K^2+M^2}\bigg]
\label{eqD5}
 \end{eqnarray} 
[the explicit calculation, $\gamma_{\alpha\alpha}^{LL} ({\bf k}) = m \partial^2 \varepsilon_L ({\bf k}) /\partial p_\alpha^2$, leads to the same result].
    The intraband conductivity spectral weight is given again by Eq.~(\ref{eqD4}), where [see Eq.~(\ref{eq11})]
\begin{equation}
n^{\rm intra}_{\alpha \alpha} = \frac{1}{V} {\sum_{L{\bf k}}}^* 
\big[ \gamma_{\alpha\alpha}^{LL} ({\bf k}) f_L({\bf k}) - \gamma_{\alpha\alpha}^{33}({\bf k}) \big],
\label{eqD6}
\end{equation}
and the total conductivity spectral weight is given by Eq.~(\ref{eqD2}), where
\begin{eqnarray}
&& \hspace{-5mm} n_{\alpha \alpha}^{\rm total} 
= \frac{1}{V} {\sum_{L{\bf k}}}^* \big[ \gamma_{\alpha \alpha}^{LL} ({\bf k};2) n_L({\bf k})- \gamma_{\alpha\alpha}^{33}({\bf k}) \big]
\nonumber \\
&& \hspace{4mm}
= -\frac{1}{V} {\sum_{L{\bf k}}}^*  \gamma_{\alpha\alpha}^{33}({\bf k}).
\label{eqD7}
\end{eqnarray}
    This result means that in the Dirac cone approximation of the 3D Dirac problem the number (\ref{eqD3}) is proportional to the integrated conductivity spectral weight measured with respect to the spectral weight of pristine compounds.
    Since it is equal to zero, the total integrated spectral weight does not change when changing the doping of conduction bands or temperature.	

%

\end{document}